# Harold Knox-Shaw and the Helwan Observatory

Jeremy Shears & Ashraf Ahmed Shaker

## Abstract

Harold Knox-Shaw (1885-1970) worked at the Helwan Observatory in Egypt from 1907 to 1924. The Observatory was equipped with a 30-inch (76 cm) reflector that was financed and constructed by the Birmingham industrialist, John Reynolds (1874-1949), to benefit from the clearer skies and more southerly latitude compared with Britain. Knox-Shaw obtained the first photograph of Halley's Comet on its 1910 perihelion passage. He also carried out morphological studies on nebulae and may have been the first to identify what later became to be known as elliptical galaxies as a distinct class of object. Photographic analysis of the variable nebula NGC 6729 in Corona Australis enabled him to conclude that the changes in brightness and shape were correlated with the light travel time from the illuminating star, R CrA.

## Introduction

By the beginning of the twentieth century fundamental changes in astronomy were well-advanced, with a move from a traditional positional and descriptive approach to the new science of astrophysics. This was driven by the development of two new tools: spectroscopy and photography. In Great Britain the chief practitioners of the science were no longer the self-taught individuals of independent wealth, the *Grand Amateurs* of the Victorian age, but increasingly they were University-trained scientists employed by professional research institutions (1). Across the Atlantic, the United States was making great strides in the development of the new astronomy. Here, in the first two decades of the twentieth century increasingly large telescopes were being built to generate astrophysical data, such as the giant reflectors on Mount Wilson which benefitted from the clear skies and southerly latitudes of California. Whilst Britain might not claim to possess the same advantages as California, it did have a different strength: a network of observatories across its colonies and protectorates around the world.

This paper describes the construction of a 30-inch (76 cm) reflector at the Helwan Observatory in Egypt during the first decade of the twentieth century and how it was used by one of its principal astronomers, Harold Knox-Shaw (1885-1970; Figure 1), in the photographic study of galaxies, comets and other objects between 1907 and 1924. After Helwan, Knox-Shaw went on to become Radcliffe Observer at Oxford and led the move of the Radcliffe Observatory from the cloud-bound UK to the vastly superior skies of South Africa. The present paper naturally focuses on the first part of Knox-Shaw's career, which hitherto has not received the attention it deserves, for what he learnt under the desert skies of Egypt influenced the rest of his career. Further information, from which to gain a broader view of his achievements, may be obtained from two obituaries written by A.D. Thackeray (1910-1978), one for the RAS (2) and one for the Astronomical Society of Southern Africa (3).

**Family life and education**

Harold Knox-Shaw was born at St. Leonards-on-Sea, Sussex, on 12 October 1885 and was the eldest of four children. His father, Charles Thomas Knox-Shaw LRCP, MRCS, MD (1865-1923; Figure 2), was a medical practitioner and was appointed as the Medical Officer of Health in Hastings in 1881. Later he converted to homeopathic medicine and became Ophthalmic Surgeon and Senior Surgeon at the London Homeopathic Hospital. He also practiced at 105 Harley Street and 19 Upper Wimpole Street, London (4). He actively promoted homeopathy through his activities as Honorary Vice President of the International Homeopathic Congress of the American Institute of Homeopathy, and as member of the British Homeopathic Association and the British Homeopathic Society.

Harold suffered a bout of polio in childhood which left him lame for life. He won an Open Scholarship to Wellington College, starting in the Lent term of 1900, where he excelled at mathematics, taking the College's Mathematics Prize four years in succession (5). In his final year he was made a prefect and a photograph of him with other College prefects is shown in Figure 3. He went up to Trinity College Cambridge in 1904 to read the Mathematical Tripos, benefitting from a Mathematical Minor Scholarship which was followed by a Major Scholarship two years later (6). He was appointed Sheepshanks Astronomical Exhibitioner and graduated Sixth Wrangler in 1907.

Harold's brother, Thomas Knox-Shaw, CBE, MC (1895 –1972), also excelled at Mathematics. He became a Fellow of Sidney Sussex College, Cambridge and Master of the College from 1945 to 1957 (7).

Harold Knox-Shaw's interest in astronomy was first aroused by a mathematics master at Wellington, Samuel Arthur Saunder (1852-1912). S.A. Saunder, also a graduate of Trinity College Cambridge, was a Fellow of the RAS and an original member of the BAA at its formation in 1890, going on to serve as President 1902-1904. Saunder owned a fine 7-inch (18 cm) refractor by Troughton & Sims, which he used mainly in selenographical studies (8). With Mary Adela Blagg (1858-1944), he set about the tedious and lengthy task of standardising lunar nomenclature and their joint *Collated List of Lunar Formations* was published in 1913 shortly after his death (9).

Another great influence on Knox-Shaw's interest is astronomy at Cambridge was Arthur Robert Hinks (1873-1945; Figure 4), under whom he worked in positional astronomy as Sheepshanks Exhibitioner. Hinks was Chief Assistant at the Cambridge Observatory from 1903 to 1913 and was responsible for the teaching of practical astronomy. He is best known for his work in determining the solar parallax conducted between 1900 and1909 and for which he was awarded the Gold Medal of the Royal Astronomical Society, also being elected a Fellow of the Royal Society. In 1913 Hinks resigned his position at Cambridge and, in a remarkable change in career direction, was appointed Assistant Secretary of the Royal Geographical Society, succeeding as Secretary two years later (10).

A year after graduation and keen to pursue research in astronomy, Knox-Shaw left Cambridge for Egypt to take up an appointment as astronomical assistant at the new Khedivial Observatory in Helwan. Britain had invaded Egypt in 1882, with the aim of securing the Suez Canal. Whilst the country continued to be ruled by the Khedive, a title largely equivalent to the English word "viceroy" (11), ultimate power resided with the British High Commissioner.

**The Khedivial Observatory and the Reynolds Telescope**

The original Khedivial Observatory was built in Abbasiya, north-east of Cairo in 1865 under the direction of the Royal astronomer, Mahmud Ahmad Hamdi al-Falaki (1815-1885) (12). The observatory had a cylindrical dome housing a 6-inch (15 cm) refractor (Figure 5). Work was conducted on both meteorology and astronomy, and the 1874 transit of Venus was observed from there (13). However, as the city of Cairo grew it began to encroach upon the observatory and interfere with observations. Thus in 1903 it was transferred to a new site at Helwan, about 25 km south of the city, at the edge of a limestone plateau on the east bank of the Nile some 150 m above sea level. Helwan attracted many tourists who visited its natural sulphur springs and river boats plying the Nile between Cairo and Luxor called in frequently. The remains of the ancient Egyptian city of Memphis lie on the opposite bank. The location was evocatively described by H.H. Turner (1861-1930) during a visit he made in 1908: (14)

"[E]very evening has a magnificent view of the Sun setting over the river among the Pyramids. The town of Helwan lies in the Nile valley below, and is laid out in the most approved modern style, with streets north and south crossed by others east to west, has several large modern hotels where there are concerts and dances in the season, and some redolent sulphur-baths". (15)

The new Khedivial Observatory, which was inaugurated on 1 January 1904, was under the control of the Survey Department of the Ministry of Finance of the Egyptian Government (16). The Survey Department was engaged in many activities, the main one being the mapping of the Mediterranean coast and the Nile valley through Egypt and Sudan, including its flood plains, as well as undertaking meteorological, geodetic and magnetic measurements, which were coordinated from its busy Headquarters at Giza, near the Pyramids. The Director-General of the Survey Department at the time was Captain Henry George Lyons (1864-1944), a geologist, who would later go on to be the first Head of the Department of Geography at Glasgow University (1909-1911) and subsequently the first Director of the Science Museum in London.

The principle building at Helwan Observatory, which still stands today, has two stories built around a courtyard, which provided welcome shade from the intense Egyptian sun, with a square tower on one side (Figure 6) housing meteorological instruments. As well as containing offices, there was a library, a meeting room, laboratories, chronometer rooms and photographic dark-rooms. There was a 35-foot (11 m) diameter dome which initially housed an 8-inch (20 cm) refractor, although it was built much larger than necessary in the hope that a sizeable telescope would be installed at some future stage. A panoramic view of the observatory location is shown in Figure 7 and a modern view of the dome in Figure 8.

In 1902 the wealthy Birmingham industrialist and philanthropist John Henry Reynolds (1874-1949; Figure 9) purchased for the sum of £80 a 30-inch (76 cm) f/4.5 mirror, one of a series which had been made by A.A. Common (1842-1903) (17). Reynolds had originally planned to build a photographic telescope incorporating the mirror for use at his residence in Harborne, south-west of Birmingham (18). However, during a visit to Egypt in 1902 he was impressed by the climate it afforded for observational astronomy and the fact that so many more southern objects could be observed compared to England. He noted: (18)

"The climate of Egypt is very suitable for astronomical photography during the greater part of the year: at times the atmosphere is so clear that stars are seen to set suddenly behind the desert horizon".

Thus Reynolds made a further visit in January 1904 with the aim of seeing whether it would be practical to establish his large reflector there. He found that the newly-built 35-foot dome at Helwan would be ideal. Moreover, he obtained the agreement of Lyons, on behalf of the Egyptian Government, to accept the telescope, which was to be a gift to the Government. As a mark of gratitude the Khedive awarded Reynolds the Order of Osmanieh (19).

Reynolds personally set about designing and building the mounting for the telescope in June 1904. The design owed much to that developed for the 60-inch (1.5 m) Mount Wilson reflector by the master telescope maker, G.W. Ritchey (1864-1945) (20). Reynolds visited Egypt again in 1905 to observe the solar eclipse in August, as well as to oversee the safe arrival of his telescope. The telescope was supplemented with a 4-inch (10 cm) Cooke photographic lens for wide field photography. Photographs of the newly mounted telescope are shown in Figures 10 and 11.

Much of 1906 was spent in making adjustments to the telescope and the first photographic exposures were made at the beginning of 1907. Writing not long after he had seen these first results, Reynolds was clearly very satisfied with his new telescope, which became known as "the Reynolds Telescope": (18)

"Optically the mirrors are of very fair figure, especially considering that the large parabolic mirror lay in a packing-case for several years; and I am glad to have been able to mount one of Dr. Common's mirrors in such a favourable situation. As far as I am aware, the Crossley reflector of the Lick Observatory is the only one in similar latitudes" (21)

Reynolds, with no formal training in astronomy, but possessing personal wealth, was perhaps one of the last of the *Grand Amateurs*. He served as RAS President between 1935 and 1937.

## Knox-Shaw begins work at Helwan

Knox-Shaw arrived at Helwan in 1907 to take up a position as assistant astronomer to B.F.E. Keeling (1880-1919) who was Superintendent (22). Keeling's responsibilities in running the Egyptian Survey Department, which he had joined in 1904, were many and varied. These included being in charge of the geodetic survey of the Nile Valley from the Mediterranean to Khartoum as well as all matters meteorological (Figure 12). Thus the Observatory was only one of many operations falling within his bailiwick. Nevertheless he was personally involved in erecting and commissioning the Reynolds Telescope and maintained a strong interest in the work of the observatory. He was frequently away on travel and with so many calls on his time the appearance of Knox-Shaw to take day-to-day charge of the Observatory must have been very welcome indeed.

The overall programme of research with the new telescope had already been set out by Reynolds: to photograph southern nebulae between the equator and -40° Declination (18). According to Keeling's report for 1908 (23), regular observational work was still rather slow in getting established, partly because of a range of mechanical teething problems which necessitated replacement parts being sent out from England and the associated delay that their shipment entailed.

H.H. Turner was invited to spend a fortnight Helwan at the end of 1908 to advise Keeling and Knox-Shaw on detailed observational plans; Turner had made a previous visit in 1905 and had seen the dome, but the telescope was not installed at the time (24). During the second visit he was fortunate enough to stay in the comfortable new rooms reserved for important visitors, in contrast to many other people passing through who stayed in tents, especially surveyors who regularly spent a few days at the Observatory whilst calibrating their instruments which they used on expeditions throughout the Middle East. Turner apparently very much enjoyed his stay, "joining the little community of University men – Cambridge mathematicians for the most part" (14) and even took to the novelty of playing golf on the Observatory's sand golf course (25). Just before leaving, Turner recorded his impression of the Observatory: (14)

"Perhaps the best view of the Helwan Observatory is from the desert just above it. The dome occupies a proud position in the front, with the other observatory buildings showing on either side, and the town of Helwan nestling in the valley in the middle distance, while the Nile and the Pyramids are away near the horizon. As I stand admiring it on the day of my regrettable departure, a train of three camels in the charge of an Arab imp winds slowly down the hill across the foreground."

Turner's visit was timely and appears to have provided the spark required, for in 1909 the first real results began to flow, including 82 photographs of nebulae (26). A new periodical was established to publish the Observatory's results: the *Bulletin of the Helwan Observatory*. But it was comet photography that was soon to project the new observatory to fame.

**The 1910 perihelion passage of Halley's Comet**

During 1909 observatories all around the world were vying with each other to be the first to detect Halley's Comet on its long anticipated return. By contrast to previous apparitions, this time astronomers had a new tool in their armoury to help them achieve an early detection: photography. As early as 1907 Reynolds had seized upon the suggestion of A.C.D. Crommelin (1865-1939), Director of the BAA Comet Section 1897-1939, to use the Helwan telescope in such a search. Thus Knox-Shaw embarked on a programme of photographing regions of the sky guided by the available ephemerides, which as it transpired were not completely reliable.

The comet was eventually found by Max Wolf in a photograph taken at the Heidelberg Observatory on the evening of 5 September 1909 and announced to the world (27). Knox-Shaw picked up the comet on 13 September (28). With new astrometry becoming available, Crommelin published an improved ephemeris on 23 September (29). Knox-Shaw re-examined his earlier plates for signs of the comet at the positions predicted by this new ephemeris and found a faint nebulous impression at the correct spot on a plate taken on 24 August. Since he did not have a measuring machine at Helwan at the time, he despatched the plate to Sir William Christie (1845-1922), Astronomer Royal, at Greenwich who had the position measured by Charles Davidson (1875-1970) (30). As Knox-Shaw noted (31), "(t)he agreement between the observed and computed positions is so small that it may be considered certain that the image measured is that of the comet". Thus whilst Wolf is correctly credited with the recovery of the comet, Knox-Shaw and Helwan can at least rightfully lay claim to obtaining the first image of the 1910 apparition 12 days earlier.

H.P. Hollis (1858-1939) was delighted that the first imprint of the comet should have been at Helwan, writing in *the English Mechanic*: (32)

"I know it is not very good manners to say 'I told you so,' but I cannot resist referring to a paragraph in my letter of Oct. 9, 1908, in which I said that I should back a dark horse - a certain observatory in North-East Africa".

Many photographs were taken from Helwan of Halley during the apparition, extending over a period of 632 days, the last being on 18 May 1911, some 394 days after perihelion (33). The majority were taken by Knox-Shaw, but others contributed too, including H.E. Hurst, an assistant at the Observatory (34). Two examples are shown in Figure 13.

Archival photographs from Helwan were re-analysed in the run-up to Halley's next apparition in 1985/6 with the aim of extracting fine details using the latest techniques in computer-based image-processing that were coming on-stream at the time. An analysis by Dan Klinglesmith (35) at the NASA-Goddard Space Flight Center of one of Knox-Shaw's plates from 25 May 1910 revealed a small anti-solar jet of material apparently being expelled from the nucleus (Figure 14). A similar feature was extracted from Helwan photographs (and from those of other observatories) in a separate analysis by Steven Larson and Zdenek Sekanina (36). The Helwan images also revealed additional fine structure near the nucleus including dust jets. Thus, more than 70 years after they were taken by Knox-Shaw, they were still revealing their secrets and such work is testament to the high quality of the original photographs.

**Observations of other comets**

Boosted by the success with Halley's Comet, Knox-Shaw resolved to photograph other comets, especially focussing on morphological studies of brighter comets and astrometry of fainter ones located at more southerly declinations (33). Noting the discrepancy between observatories in the reported magnitudes of Comet Halley when faint, Knox-Shaw devised a new method of comparing the brightness of the comet with field stars on photographs which he applied to other comets (37).

Knox-Shaw achieved a personal record on 25 October 1911 when he managed to photograph five comets in one night: C/1991 O1 Brooks, 19P/Borelly, C/1911 S2 Quenisset, C/1911 S3 Beljawski and 14P/Wolf (38). Knox-Shaw and Alexandre Schaumasse (1882-1958), of Nice Observatory, had independently recovered 19P/Borelly on it second observed apparition in late September 1911 (39). This turned out to be a particularly favourable apparition and a photograph is shown in Figure 15.

Naturally, many other comets were tracked from Helwan over the years, including 28P/Neujmin in 1913 (40) and C/1915 C1 Mellish which showed two condensations in the nucleus (41).

**Morphological studies of southern nebulae**

Reynolds' main wish for the Helwan Observatory was that the telescope should be used to photograph southern nebulae. Even before the telescope was available for use, Knox-Shaw gained experience in measuring the inclinations of spiral galaxies in photographs taken by the late Isaac Roberts (1829-1904) (42). His own photographic work began in earnest in 1909, embarking on nebulae listed in the NGC (43). Generally he employed exposures of 1 to 2 hours. He measured the position and dimensions of the object from the plates and gave a description, mainly following the nomenclature developed by William Herschel. At the time the work began, whilst different categories of nebulae were recognised such as gaseous nebulae, planetary nebulae and spiral nebulae, it was not known that some nebulae are located within our galaxy and that some, in particular the spiral nebulae, are external galaxies, or "island universes". Knox-Shaw found that many objects, which looked like nebulous stars on photographs, some of which had been classified by Herschel as globular clusters, constituted a class of object which he termed "globular nebulae". These are now known to be elliptical galaxies and Knox-Shaw may have been the first to recognise them as a distinct group. During his work he encountered numerous previously unknown nebulae, many of which turned out to be elliptical galaxies. A total of 322 nebulae were photographed in the course of the project between 1909 and 1922. The work was published in a series of papers in the *Bulletin of the Helwan Observatory* (44) and some examples of objects are shown in Figure 16.

Reynolds was especially interested in galaxy morphology and classification and he was therefore delighted with the photographs coming from Helwan which helped him greatly in these studies. In 1920 he published in the *Monthly Notices of the Royal Astronomical Society* (45) a classification scheme with seven different categories, largely based on the degree of condensation of the nebulosity measured from Helwan plates of representative galaxies. His classification was very similar to that proposed some 6 years later by Edwin Hubble (1889-1953), of Sa, Sb, Sc etc. Hubble was certainly aware of Reynolds' classification work – they corresponded regularly and they met on several occasions in England. Conclusive evidence that Hubble knew of Reynolds' classification scheme has recently been brought to light by research conducted by David Block (University of the Witwatersrand, South Africa) and Kenneth Freeman (Australian National University) (46). They found a memo in the RAS archives from Hubble to Vesto Slipher (1875-1969), dated July 1923, entitled "*The Classification of Nebulae*" stating:

"The published suggestions of J.H. Reynolds are thoroughly sound…Reynolds introduced the term amorphous, emphasizing the unresolvable character of much of the nebulosity in non-galactic spirals…Reynolds (ref. 17) has formulated seven classes of true spirals…the first five classes represent a series with increasing degree of condensation in the amorphous matrix of the outer arms…."

The "reference 17" which Hubble mentions in his memo is none other than Reynolds' 1920 *MNRAS* paper. In spite of this, Hubble's 1926 paper on his own classification scheme makes no reference to Reynolds' pioneering work and history has credited Hubble alone.

**The Variable Nebula in Corona Australis**

In 1861 Julius Schmidt (1825-1884) of the Athens Observatory discovered a trio of nebulae whilst observing the globular cluster NGC 6723 in Corona Australis: NGC 6726-7 and 6729 (47) (Figure 17). The latter is associated with the variable star R CrA which is located at its northwest end and which he also discovered. He suspected the nebula to be of variable brightness. The nebula was independently discovered by Albert Marth (1828-1897) using William Lassell's (1799-1880) telescope in Malta in 1864 (48).

Knox-Shaw's photographs of the object with the Reynolds telescope during the summer of 1911 showed conclusively that the nebula was variable in both brightness and apparent shape (49). He continued monitoring it into the 1920s and at times found noticeable variations in the nebula from one day to the next. He classified the appearance of the nebula into 8 categories, which he called Type A to H, some of which are illustrated in Figure 18. He concluded that the changes in brightness and shape were correlated with the light travel time from R CrA, the result of a so-called light echo (50). In the 1980s, studies (51) of the long-term optical behaviour showed that the changing appearance of the nebula is also caused by the shadowing effects of clouds which are close to R CrA, possibly within 1 AU (52).

Knox-Shaw was also interested in the "dark spaces in the Milky Way" in the vicinity of R CrA, and elsewhere, that had also been pointed out by E.E. Barnard (1857-1923) as a result of his wide-field photographic survey. The cause of these dark spaces had been much debated and Knox-Shaw was of the opinion, as was Barnard, that these were due to obscuring matter rather than actually being star-poor regions, or "holes" in the Milky Way (53). A modern image of the region of R CrA clearly shows the dark nebula B157 (Figure 17 (b)).

In 1916 Hubble announced his discovery of the variability of another nebula, NGC 2261 in Monoceros, now commonly known as *Hubble's Variable Nebula* (54). The object is associated with the variable star, R Mon. Recognising its apparent similarity to the nebula on CrA, Knox-Shaw began to photograph NGC 2261 in December 1916 and continued to March 1920, confirming changes in its form and brightness, some examples of which are shown in Figure 19.

### Jupiter's eighth moon

Through the application of photography, several discoveries of new moons of Jupiter were made during the course of the first decade of the twentieth century. Jupiter VI (Himalia) was discovered in 1904, J VII (Elara) in 1905 and J VIII (Pasiphae (55)) was first spotted by Philibert Melotte (1880 - 1961) (56) on a plate taken at the Royal Observatory Greenwich on the night of 28 February 1908. Inspection of earlier plates showed it to be present as far back as 27 January. Following the announcement of the discovery it received the provisional designation 1908 CJ, as it was not immediately certain whether it was an asteroid or a moon of Jupiter. The recognition of the latter case came by April 10 (57).

Knox-Shaw, often assisted by T.L. Eckersley, obtained astrometry of the moon from plates taken with the Reynolds telescope during the oppositions of 1911 to 1916 (58). These data were used to update the orbital ephemeris. However, no observations of J VIII were obtained by any observatory during the 1919 to 1921 oppositions (Knox-Shaw made unsuccessful attempts in 1921) and the moon was regarded as lost.

However, it was recovered again at Helwan on 23 February 1922 and a long series of observations were obtained lasting until 20 May (59). It was also picked up on 5 nights in April by George van Biesbroeck (1880-1974) and Otto Struve (1897-1963) using the 24-inch (61 cm) reflector at Yerkes. The new observations from both observatories, combined with previous data, allowed John Jackson (1887-1958) at Greenwich to compute a more accurate orbit (60).

## Upgrading the Reynolds telescope

Knox-Shaw and Reynolds continued to make improvements to the 30-inch telescope as their experience with it grew. In 1910 the pier was raised by 2-foot to allow the telescope to reach slightly lower declinations (61). At the same time a 5-inch (13 cm) refractor was added as a photographic guide telescope (62). The following year a new clock drive with electrical control was fitted, which facilitated long-exposure photography. A new Newtonian flat was also installed, which had been made by G.W. Ritchey (63). In January 1914 an improved driving worm and sector were installed, the original having been re-cut by the firm of Sir Howard Grubb at Newcastle-upon-Tyne (64).

A constant problem encountered was rapid deterioration of the reflectivity of the mirror each time it was re-silvered. This was partly due to dust deposition from the desert air and partly due to chemical tarnishing, thought to be related to the proximity of the sulphur springs which were about 3 km away. It was noted that spots began to form on the mirror within a few weeks of silvering and after about 6 months they were so numerous that the mirror needed re-silvering. Knox-Shaw and his colleague S.H. Trimen, of the Egyptian Survey Department Laboratories, made several attempts in 1910 and 1911 to give the mirror a protective coating of collodion, having read a report that this had been done at the Meudon Observatory, but unfortunately this was not successful as it spoiled the definition and had to be removed (65).

However, the most significant improvement to the telescope was the installation of a new 30-inch primary. Whilst the Common mirror was good, it was recognised that the figure was not perfect. The wealthy philanthropist, William Waldorf Astor (1848-1919), made funds available for a replacement mirror in 1912 and an order was placed with Ritchey. Ritchey, of course, also made the mirrors for both the 60-inch (1.5 m) and 100-inch (2.5 m) telescopes on Mount Wilson. Before it was shipped the mirror was extensively tested at the request of H.H. Turner by W.S. Adams (1876-1946) of the Mount Wilson Observatory (66). Knox-Shaw subjected the original Common mirror to the same tests (67), finding it to be a little under-corrected and slightly astigmatic. The new mirror was installed during October and November 1914 (68) and was found to yield superior images both visually and photographically, leading Knox-Shaw to re-photograph many nebulae over the next few years.

As part of the upgrading project Reynolds made a completely new polar axis which allowed the telescope to reach a wider range of declinations. This was installed at about the same time as the Ritchey mirror.

The story of the Common mirror did not end there. Reynolds had it sent back to Britain where he installed it in place of the 28-inch mirror in his telescope at Harborne. He continued to use the "new" 30-inch reflector for several years. He obtained a quote from Sir Howard Grubb, Parsons & Co. in 1927 (69) to either refigure the 30-inch mirror (at a cost of £360) or to provide an entirely new mirror (£600), but the authors have not been able to ascertain whether he acted upon the quote.

Reynolds donated the Harborne telescope to the Mount Stromlo Observatory in Australia where it was set up in 1929. This instrument was also referred to as the Reynolds telescope and until the 1950s it was the largest working telescope in the Southern Hemisphere. The instrument was employed by Gérard de Vaucouleurs (1918 –1995) between 1952 and 1955 to conduct a photographic survey of bright southern galaxies, which was effectively an extension of the Helwan Survey (70). The telescope saw active service until it was destroyed by the devastating bush fires of 2003.

**Career progression**

In 1913 the scientific branches of the Egyptian Survey Department, including the Helwan Observatory, were restructured to form the Physical Service with Keeling as Director. As a result Knox-Shaw was promoted to the position of Superintendent of the Observatory, a position he held until he left Egypt in 1924. With the commencement of the First World War, Keeling returned to Europe in December 1914 where he received a Commission in the Royal Engineers and served on the Western Front, reaching the rank of Lieutenant-Colonel. Knox-Shaw remained in Egypt during the war where he was engaged in welfare work amongst British troops. His work was recognised by the award of the Order of the Nile, Class 4. Knox-Shaw took on more responsibilities as a consequence of Keeling's absence. He endeavoured to keep the observatory operating during the war, although pressure of work elsewhere meant that that the Reynolds telescope was idle for 1917, 1918 and most of 1919. In 1916 he sailed to Britain for a period of home leave, a voyage not without considering the vulnerability of merchant ships to German U-Boats. He was due to speak at the June RAS meeting, but did not arrive in time since a more tortuous route than normal was taken. He did, however, mange to address the June 1916 BAA meeting about the work of the Helwan Observatory (71).

Knox-Shaw was appointed Director of the Meteorological Services of Egypt and the Sudan in 1918, a position he held concurrently with his Observatory role. Meanwhile Keeling was demobilised in the spring of 1919 and returned to Egypt in April as head of the Survey Department and chairman of the newly formed Board of Cotton

Research, but he died suddenly on 25 September 1919 (22). With the return of peace Knox-Shaw arranged a complete overhaul of the Reynolds telescope in November 1919 and photographic work resumed later that month.

**Chronic understaffing**

Scientific work at the Helwan Observatory was often hindered due to a lack of staff available to operate the telescope. Knox-Shaw also found that he had other responsibilities beyond the Observatory and was often drawn into surveying work. Whilst he did not usually join the surveying expeditions himself, he was involved in analysing and publishing the data. Figure 20 shows a map of northern Egypt and the Mediterranean coast he drew based on field data obtained by E.B.H. Wade, a colleague in the Survey Department, and Lt. E.R. Pratt of the Royal Artillery. Wade, in turn, also assisted in the Observatory occasionally – it was common for colleagues to help each other out in this way.

Another disruption to astronomical research was when Knox-Shaw went on home leave as there was often no one to take his place; for example, no observations were made in July to September 1910 due to his absence (61).

H.H. Turner recognised the chronic understaffing and suggested that suitably skilled volunteers would be welcome as assistants at the Observatory (72). One who responded was the Director of the BAA Photographic Section, F.W. Longbottom (1850-1933) (73). In late October 1911 Longbottom, a retired hop-merchant from Chester, set sail for Egypt on board P&O's *SS Mongolia* where he planned to overwinter for 4½ months. He worked alongside Knox-Shaw and participated in the Observatory's photographic programme, where his experience in astronomical photography was especially helpful, exposing and developing plates and preparing them for publication (74). Longbottom donated a hand-held spectroscope to the observatory which Knox-Shaw sometimes used to confirm the identity of gaseous nebulae. Longbottom clearly enjoyed his time at Helwan and he developed a continuing friendship with both Knox-Shaw and Reynolds as a result. He described his experiences in talk at the Liverpool Astronomical Society under the title of "An Astronomical Holiday" (75).

Several colleagues who assisted Knox-Shaw over the years went on to become well-known in their own scientific careers. One example was Harold Edwin Hurst (1880-1978; Figure 21), who, as noted before, occasionally assisted Knox-Shaw at the Observatory, notably in photographing Halley's Comet during the 1910 apparition. His main responsibilities in the Survey Department, which he joined in 1906, were in connexion with meteorology and hydrology. He went on to become an eminent hydrologist and continued to have a long association with Egypt throughout his career during which he did much work on the hydrology and flooding of the Nile, becoming known as the "Father of the Nile" (76).

Another example was T.L. Eckersley (1886-1959), a graduate of Trinity College Cambridge, who joined the Helwan Observatory in 1913 to assist in magnetic survey work as well as photographic work with the Reynolds telescope (77). He also carried out measurements of solar radiation in an attempt to determine the solar constant, but it transpired that the Helwan site, at a very modest altitude, was not really suitable (78). As described previously, one of his first projects was to help Knox-Shaw with photographing J VIII in 1913 until he left Egypt for military service in November 1915. Later, Eckersley became interested in radio-wave propagation and he was one of the first to obtain evidence of the presence of the ionosphere by the reflection of radio waves. He pursued a career with the Marconi Company at their Research Laboratories, becoming a Fellow of the Royal Society in 1938 (79).

Eckersley was succeeded by E.W. Bliss who also continued to assist with the general photographic work as well as the unsuccessful search for J VIII in 1921 (80).

In 1919, in an effort to provide Knox-Shaw with more support to recover from the slow-down due to the First World War, the position of Chief Assistant at the Observatory was created and filled by Christopher Clive Langton Gregory (1892-1964) (81). C.C.L. Gregory had graduated from Cambridge with a degree in Mathematics in 1915 and started research at Imperial College London, although this was soon interrupted by war service. At Helwan he became involved in the nebula photography project. Whilst Knox-Shaw was glad to have an assistant whose primary responsibility was astronomy, this was tempered by Knox-Shaw having to spend even more of his own time on other activities, especially meteorology (82). After Gregory returned to England in 1921 he took up the post of Lecturer in Astronomy at University College, London (83). Having a strong interest in observational astronomy, developed during his time at Helwan, he was determined that the College should have an observatory of its own. Subsequently the University of London Observatory was established at Mill Hill and he was appointed as its Wilson Observer in 1928 (Figure 22).

The well-known BAA personality W.H. Steavenson (1894-1975) visited Helwan in 1920, whilst he was serving in Egypt for six months with the Royal Army Medical Corps, and observed Mars and Mercury from the observatory. He might have encouraged Knox-Shaw, who was already an FRAS, to become a member of the BAA as he joined the Association not long after, with Steavenson as his proposer (84).

**Egyptian independence and Knox-Shaw's later years at Helwan**

The period after the end of the First World War was a turbulent one in Egypt. In 1914 as a result of the onset of war with the Ottoman Empire, with which Egypt had an historical association (85), Great Britain declared a Protectorate over Egypt and deposed the Khedive. He was replaced with a family member, Faud, who was made Sultan of Egypt. However, with the end of the war and with the world map being redrawn, a growing independence movement developed. A group known as the Wafd Delegation attended the Paris Peace Conference in 1919 to demand Egypt's full independence. Members of the group were arrested and swiftly deported, leading to mass demonstrations and violence in Cairo during March and April 1919. In November 1919 the British sent Lord Milner to Egypt to attempt to resolve the situation and calm was restored temporarily. Discussions took place between the British Foreign Secretary, Lord Curzon, and delegates from Egypt in 1920 and 1921, but they could not agree terms of control of the Suez Canal Zone, over which Britain was determined to maintain control. In December 1921, Britain imposed martial law as violence erupted once again. Following a suggestion by the High Commissioner Lord Allenby, Britain unilaterally declared Egyptian independence on 28 February 1922. However Britain still maintained strong links with the Egyptian government and retained control of the Canal Zone.

Given the proximity to Cairo, the scientists at Helwan were well aware of the political developments and the demonstrations and violence that erupted in the city from time to time. Knox-Shaw made frequent references to the turbulent situation in his letters to Reynolds. The two had become firm friends since Reynolds made numerous visits to Helwan during which he participated in the work of the Observatory (86). Knox-Shaw also visited Reynolds' residence in Harborne and knew his family well. In early November 1921 Knox-Shaw wrote to Reynolds about plans for his next visit to Helwan, saying he was about to re-silver the 30-inch mirror in preparation. It was shortly after this that violence broke out again in Cairo, prompting him to write to Reynolds, reassuring him about the situation (part of the letter is shown in Figure 23): (87)

"There has been some rioting in Cairo since I last wrote but things seemed to have settled down again and if they don't burst out again I see no reason why you should put off your visit. It has been perfectly quiet here and there have been tourists up the river and Cooks steamers are running……We do not expect anything much more in the way of trouble. It has been nothing to compare with the 1919 events"

He goes on to comment about his other responsibilities at Helwan which were taking him away from astronomy:

"I am afraid that very shortly I shall have to take up my meteorological work again. I shall hate it, but as usual in this wicked world one has to do what one does not want!"

In the event, Reynolds did put off his visit and the political situation remained tense for some time. Tourists were also staying away. Nevertheless W.H. Pickering (1858-1938) visited the Observatory in mid February 1922, following a trip to Luxor, and was hosted by Knox-Shaw (88).

During 1922, Knox-Shaw oversaw the installation of a new radio receiver at the Observatory (Figure 24) which enabled more precise time measurements to be made via wireless telegraphy (89). Combining wireless time signals with stellar transit timings measured with the Observatory's Brunner Transit Circle (Figure 25), allowed the observatory's longitude to be determined more accurately than hitherto (90).

As 1922 progressed and the independent Egyptian government was established, Knox-Shaw became increasingly frustrated about the lack of time he had available for observational astronomy, especially as by then Gregory had already been gone for more than a year and there were no definite plans to appoint a replacement. One of the stumbling blocks was the insistence of the new Egyptian government that any new astronomer should be an Egyptian national (91). Knox-Shaw confided to Reynolds in August: (92)

"I have had more day work than I can get through…I am spending half my time in Cairo and half here, but my time here is mostly taken up with the routine work of the observatory: magnetism, seismology etc…..I put Hurst to go and see whether our Under Secretary of State, now an Egyptian, would consent to the appointment of an astronomer for a period of two years. We pointed out that there were no Egyptians qualified at the moment and none likely to want to be astronomers. The latter point did not worry him. He said a man was to be chosen and sent to England to be trained as an astronomer whether he wanted to or not and absolutely refused to allow an Englishman to be appointed in the meantime."

Eventually the Egyptian Educational Mission contacted H.F. Newall (1857-1944), Professor of Astrophysics at Cambridge and Director of the Cambridge Observatories, to enquire whether he could "take on a young Egyptian, selected by the Egyptian Govt, and give him opportunities for study in Physical Astronomy" (93).

Clearly Knox-Shaw felt that the time had come to look for new employment opportunities elsewhere. In early 1924 he began to have discussions about taking up the position of Radcliffe Observer, in charge of the Radcliffe Observatory at Oxford. The initial response of the Radcliffe Trustees was encouraging and they suggested that he make enquiries with the Egyptian authorities about being released from his position at Helwan before the end of the statutory 6 months notice period. There was some urgency on the part of the Trustees as the previous Radcliffe Observer, A.A. Rambaut (1859-1923), had died the previous October and they had enlisted the services of J.L.E. Dreyer (1852-1926), in retirement at Oxford, as their adviser in the interim. However the Egyptian authorities would not accede to Knox-Shaw's request and consequently he feared that the job in Oxford might slip through his hands (94). Happily he received news from Oxford in early May 1924 that the position was his and immediately he began to make plans to leave Egypt. However after 17 years at Helwan he had mixed feelings about his impending departure, commenting to Reynolds: (95)  "It is going to be perfectly horrid leaving here, but I expect that once the change has been made I shall feel that I have done the right thing". After much negotiating, he managed to secure an early release from his responsibilities at Helwan noting (96) that "[T]hey do everything these days they can do to annoy an Englishman in the Service".

Knox-Shaw arrived in England in August 1924 and proceeded directly to Oxford and the Radcliffe Observatory to take up his new position. Thus the Egyptian chapter of his life had drawn to a close. In 1926 the Egyptian Government awarded him the Order of the Nile, Class 3, in recognition of his astronomical work at Helwan.

Knox-Shaw's time at Helwan was generally happy and fulfilling in spite of the frustrations about the chronic lack of manpower. A photograph of him at the observatory taken towards the end of his tenure there is shown in Figure 26. In later years he told anecdotes about life there. His son Peter recalls a couple: (97)

"Like many colonials of his period, HKS delighted in the vagaries of second-language English. So we often heard about the caretaker at Helwan who suddenly announced, 'my wife was subtracted by my friend in the night, and O my God I am annoyed', or who put in a written application for leave on the grounds that his grandmother had died that morning and he had toothache.

One frequent summer resorter to Helwan was the composer Ethel Smyth (98) with whom my father often played tennis. She hated the long dress that was *de rigueur* and would disappear behind a bush, between sets, to remove layer after layer of clothing until she emerged at last in something as skimpy as the shorts of Billy Jean".

**The post-Knox-Shaw era at Helwan**

Knox-Shaw was succeeded as Director of the Helwan Observatory by P.A. Curry, who had already worked there for some time. However, Curry's main responsibilities lay in meteorology, consequently, with manpower even more constrained, astronomical work took a back seat and very few observations were made with the 30-inch telescope in 1925 and 1926. Finally a new Resident Astronomer was appointed in 1926, Dr Mohammed Reda Madwar (1893-1973; Figure 27) (99). Madwar had attended French schools in Egypt and Paris. He graduated BSc. at Edinburgh University in 1917 and remained in Britain for a further two years working as an engineer. He returned to Egypt in 1919 and taking a position in the Ministry of Public Works for four years. He was then awarded a scholarship which enabled him to return to Edinburgh where he undertook research in astronomy, graduating PhD in 1926, joining the BAA the same year (100). It was at this point that he took up the appointment at Helwan, holding the post of Resident Astronomer until 1934 when he succeeded Curry, becoming the first Egyptian Director of the Helwan Observatory. In 1936 Madwar was appointed Professor of Astronomy at Cairo University, which post he held alongside his Helwan directorship.

Madwar's first task at Helwan was to bring the Reynolds telescope back into action which he had succeeded in doing by the end of November 1926. Knox-Shaw took leave of absence of a few weeks from Oxford to assist with the re-commissioning, starting in December and spending New Year there. Writing to Reynolds (101) shortly before leaving for home in the middle of January 1927 he said that he had "had an awfully good time and do not want to return at all, but I must as I shall have overstayed my leave by a day as it is". Madwar submitted his first monthly report to Reynolds on the use of the telescope in February 1927 (102). Madwar continued the work of photographing galaxies with the Reynolds telescope that Knox-Shaw had started.

The expansion of industry in the neighbourhood of the observatory after the Second World War once again necessitated it be moved, this time to the desert site of Kottamia, 80 km north-east of Helwan. Madwar led the site testing and eventually a new dome housing a 74-inch (1.9 m) telescope constructed by Grubb Parsons was opened in 1963 and continues to be used in research today. The Helwan Observatory and the Reynolds telescope still stand and are part of the National Research Institute of Astronomy and Geophysics (103). The Reynolds telescope is no longer used for scientific research, but is open for inspection by visitors, being part of a museum where one of the present authors (AAS) works.

**The Radcliffe Observer and a move to South Africa**

Having grown accustomed to the clear skies of Egypt, with the ability to see southerly objects, returning to cloudy Oxford was a frustrating experience for Knox-Shaw. Gone were the days when he could follow a setting star to the horizon before it was suddenly extinguished as Reynolds had found during his first visit to Egypt. He was convinced that the Oxford climate was most unsuitable for observational astronomy with large telescopes, enjoying fewer than 30 nights per year which were acceptable for high quality photography (104). By 1928 he had begun seeking support to move the Radcliffe Observatory to a better location overseas, with a preference for South Africa (2) (3). Amidst much opposition to the move from certain influential quarters at the University, which ultimately involved litigation, Knox-Shaw eventually won the day, with support from an array of internationally recognised astronomers who supported his cause. An order for a 74-inch (1.9 m) telescope was placed at Grubb Parsons in 1935. The design of the observatory building incorporated features requested by Knox-Shaw to minimise diurnal temperature variations, based on his experience at Helwan (105). By 1939 Knox-Shaw was in residence at the new Observatory in Pretoria. The onset of the Second World War inevitably delayed the construction of the telescope, but the mirror arrived in 1948 and the telescope eventually became operational in 1951, shortly after Knox-Shaw's retirement.

Knox-Shaw evidently had great personal charm and warmth which easily won him friends. He also had a great tenacity of purpose and meticulous attention to detail (2). These qualities stood him in good stead during the difficulties surrounding the move of the Radcliffe Observatory from Oxford to South Africa.

Knox-Shaw married Maisie Weir of Pretoria and their son, Peter, was born in 1944. He made one further visit to Helwan in 1965, during a visit to England. There he met Professor Madwar and other staff members. Peter Knox-Shaw recalls that the reunions on both sides were very emotional (97).

Knox-Shaw spent the rest of his life in South Africa and passed away in Cape Town on 11 April 1970, following a stroke (2).

## Acknowledgements

So many people have been very helpful to us in this research and we are grateful for their generosity. We especially acknowledge Peter Hingley, RAS Librarian, whose enthusiastic encouragement is greatly appreciated. Peter allowed JS access to RAS archival material on Reynolds. He also tracked down photographs of the Helwan telescope in the RAS archives and gave us permission to use them in this paper. Sadly Peter passed away just as the first draft of the paper was nearing completion. He will be greatly missed. Peter Knox-Shaw, Harold's son, has been extremely helpful in providing background information about his father's time in Helwan, as well as providing helpful comments on an early draft of the paper. He gave us permission to publish copies of his father's photographs of Halley's and Borelly's Comets held at the Science & Society Picture Library and the Library's Sophia Brothers kindly organised scans to be made of the original plates. Peter Knox-Shaw also allowed us to use copies of his father's personal photographs taken at Helwan using a stereoscopic camera and which are published here for the first time. Anne Charles, of the South African Astronomical Observatory and the Friends of the Cape Town Observatory, is in the process of cleaning and preserving the original photographs and she kindly brought forward the scanning of some to enable them to be included in this paper. The original photographs, of which only are few are used here, are in the care of Green Templeton College, Oxford. Joy Wheeler of the Royal Geographical Society's Picture Library gave permission use the image of Arthur Robert Hinks. Remus Chua, of Singastro, Singapore, allowed us to use his beautiful shot of the region around the variable nebula in CrA and prepared the image for publication. Sylvain Cazalet, *Homéopathe International*, allowed us to use the photograph of Charles Knox-Shaw and Sue Young, of Sue Young Histories, kindly provided background information about Charles Knox-Shaw. Dan Klinglesmith, of the Magdalena Ridge Observatory at the New Mexico Institute of Mining and Technology, gave permission to use photographs from his paper in which he subjected Knox-Shaw's photographs of Halley's Comet to modern image processing techniques. Ken Freeman, Research School of Astronomy & Astrophysics at the Australian National University, provided information about the fate of the "other" Reynolds telescope, i.e. the one which Reynolds installed at Harborne and subsequently gave to the Mount Stromlo Observatory. Chris Potter, Secretary of Wellington College's Old Wellingtonian Society, kindly searched the College archives and provided information about Knox-Shaw's school days, including the photograph of him as a prefect, school registers and yearbook entries. Ian Howarth, University College

London and University of London Observatory, gave permission to use the photograph of C.C.L. Gregory. Vintage car expert Michael Worthington Williams kindly identified the Buick motor car that is shown here, with Knox-Shaw in the driving seat. Finally we thank our referees for their helpful comments and suggestions.

This research made use of the NASA/Smithsonian Astrophysics Data System, the Archives of the National Research Institute of Astronomy and Geophysics, Egypt and the Archives of the RAS.

**Addresses**

JS: "Pemberton", School Lane, Bunbury, Tarporley, Cheshire, CW6 9NR, UK [bunburyobservatory@hotmail.com]

AAS: National Research Institute of Astronomy and Geophysics, Helwan (11421), Cairo, Egypt [shaker@nriag.sci.eg]

**Notes and references**

10. Hinks' obituary, written by Henry Norris Russell (1877-1957) who had worked with Hinks at Cambridge between 1903 and 1905, is at: Norris H.N., MNRAS, 106, 30-31 (1946). A more detailed obituary is at: Smart W.M., Obs., 66, 89-9 (1945).

11. The term Khedive was first used early in the nineteenth century by Muhammad Ali Pasha, the Wāli (Governor) of Egypt and Sudan, and vassal of the Ottoman Empire.

12. Al-Falaki is Arabic for "the astronomer", which became his family name.

13. Browne O., MNRAS, 36, 101 (1876).

14. Turner H.H., Obs., 32, 110-114 (1909).

15. Turner even likened the atmosphere of Helwan to that of the English seaside resort of Southport, with its parks and playgrounds for children, and he commented that he half expected to see the expanse of the Irish Sea around the next corner!

16. Keeling B.F.E., HelOB, 1, 1-2 (1911).

17. Common had made several 30-inch mirrors of the same focal length for astrographic applications. One was installed in the Thompson Photographic Reflector at the Royal Observatory Greenwich in 1896. The precise focal length of the Helwan mirror was determined by Knox-Shaw to be 3495 mm and the plate scale at the Newtonian focus was 1 mm = 59 arcsec; Knox-Shaw H., MNRAS, 71, 573-577 (1911).

18. Reynolds J.H., MNRAS, 67, 447-449 (1907).

19. Obituary of J.H. Reynolds: Knox-Shaw H., Obs., 70, 30-31 (1950).

20. Dr. Ron Maddison has Reynold's personal copy of Ritchey's paper on the design of the Mount Wilson telescope, published in 1904, and this contains Reynold's handwritten annotations. Maddison R., The Antiquarian Astronomer, issue 5, 36-40 (2011).

21. The 36-inch (91 cm) Crossley came into operation at Lick in 1896. Ritchey's 60-inch (1.5 m) Crossley reflector on Mount Wilson, only ~4 degrees north of the latitude of Helwan, was to see first light in December 1908.

22. Further details about Keeling may be found in his obituaries: Lyons H.G., MNRAS, 80, 347-348 (1920) and Lyons H.G., Nature, 104, 317-318 (1919).

23. Keeling B.F.E., MNRAS, 69, 284 (1909).

24. Turner H.H., Obs., 28, 361-366 (1905).

25. According to at least one of the authors, there are very few things that could make a game of golf palatable and sand certainly isn't one of them.

26. Keeling B.F.E., MNRAS, 70, 333 (1910).

27. Wolf M., AN, 182, 179 (1909).

28. AN, 182, 227 (1909).

83. In the spring of 1921, when his assignment at Helwan was nearing an end, Gregory wrote to Reynolds enquiring if he might know of any suitable positions in astronomy (letter dated 7 April 1921; Archives of the RAS).

84. JBAA, 32, 159 (1922).

85. Even though Egypt was largely under the British sphere of influence since Britain's invasion in 1882, it was still nominally an autonomous vassal state of the Ottoman Empire. Britain's aim in 1914 was to "clarify" the situation.

86. A non-astronomical highlight of these visits was when Reynolds, an accomplished musician, played the organ in the English church in Helwan; Knox-Shaw H., Obs., 70, 30-31 (1950).

87. Letter from Knox-Shaw to Reynolds, 30 December 1921. RAS Archives.

88. Letter from Knox-Shaw to Reynolds, 5 February 1922. RAS Archives.

89. Investigations were also made into small errors in timings determined via wireless time signals that had been reported at other observatories, but the results were inconclusive. Knox-Shaw H., HelOB, 28, 17-59 (1923); HelOB, 31, 87-128 (1924).

90. Knox-Shaw H., HelOB, 25, 269-279 (1922).

91. Gregory returned to England in July 1921.

92. Letter from Knox-Shaw to Reynolds, 13 August 1922. RAS Archives.

93. Letter from H.F. Newall to Reynolds, 22 September 1923. RAS Archives.

94. Letter from Knox-Shaw to Reynolds, 12 April 1924. RAS Archives.

95. Letter from Knox-Shaw to Reynolds, 10 May 1924. RAS Archives.

96. Letter from Knox-Shaw to Reynolds, 26 May 1924. RAS Archives.

97. Knox-Shaw P., Personal communication (2012).

98. Ethel Smith (1858-1928) was an English composer, a leader of the women's suffrage movement and a close friend of Emmeline Pankhurst.

99. An outline of Madwar's life may be read in an online article by O'Connor J.J. & Robertson E.F. (2007) http://www-history.mcs.st-and.ac.uk/Biographies/Madwar.html.

100. Madwar was elected to the BAA on 27 October 1926. His proposers were P.J. Melotte and A.C.D. Crommelin. JBAA, 36, 319 (1926).

101. Letter from Knox-Shaw to Reynolds, 16 January 1927. RAS Archives.

102. All Directors of the Helwan Observatory were obliged to submit a monthly report on the use of the telescope. They usually report the number of nights the telescope was used, the number of plates exposed and a brief summary of the objects photographed and the results obtained. Many of these reports are in the RAS Archives.

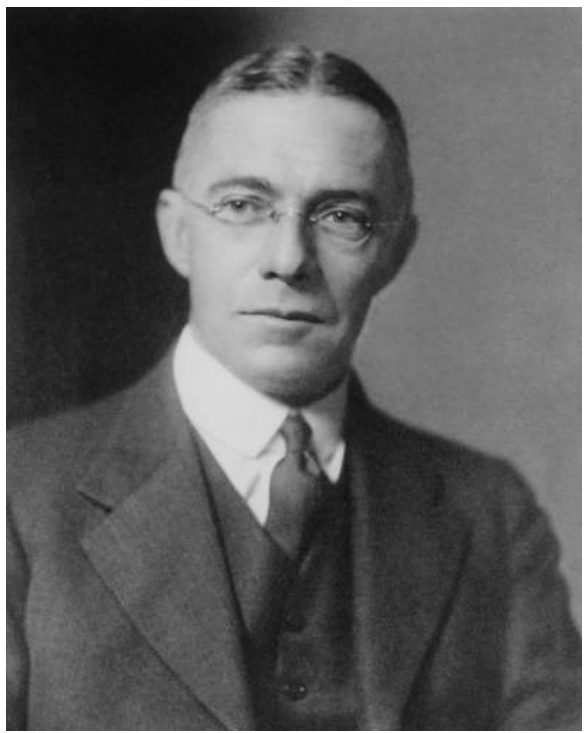

Figure 1: Harold Knox-Shaw (1885-1970)

(Royal Astronomical Society (106))

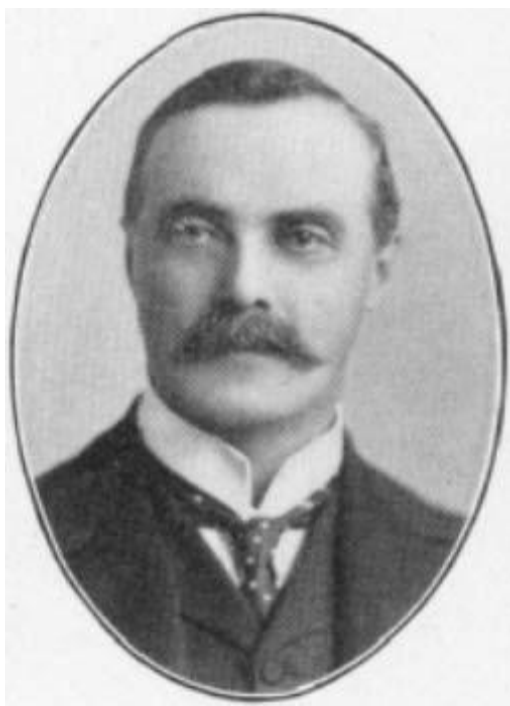

Figure 2: Charles Thomas Knox-Shaw LRCP, MRCS, MD (1865-1923)

(image from the collection of Sylvain Cazalet, *Homéopathe International,* and used with permission (107))

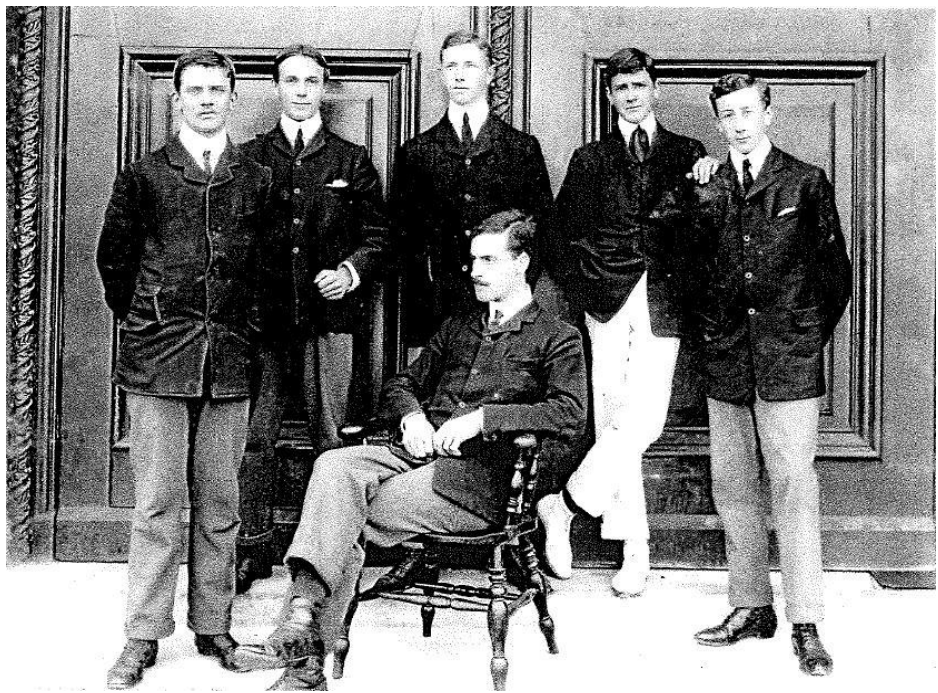

Figure 3: Wellington College prefects in 1904

Knox-Shaw is standing, second from left. The head boy is seated

(image courtesy of Wellington College)

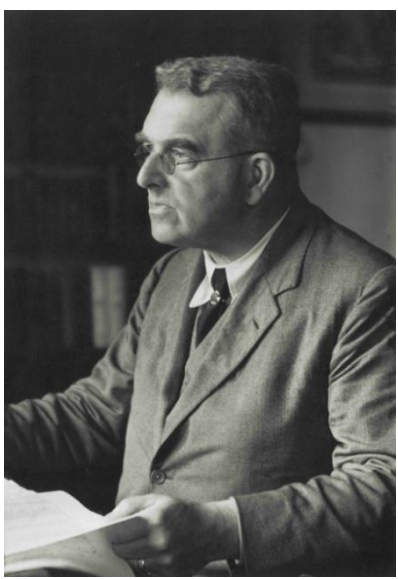

Figure 4: Arthur Robert Hinks, CBE, MA, FRS (1873-1945)

(image courtesy of Royal Geographical Society)

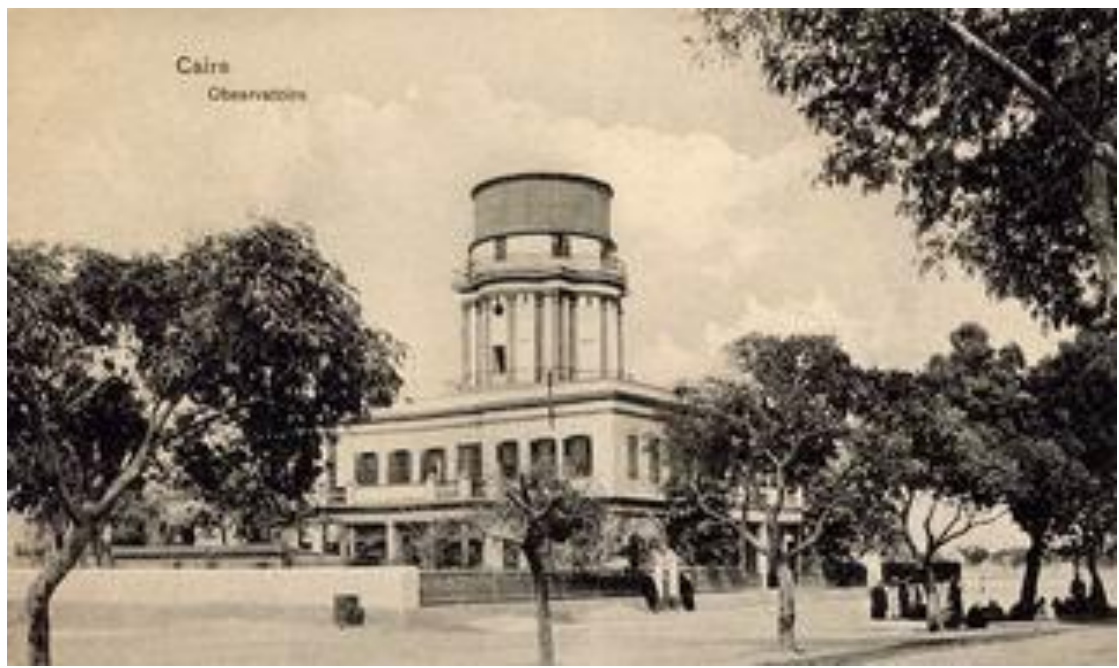

Figure 5: The original Khedivial Observatory at Abbasiya on a 1906 postcard. The building remained until at least 1936. Today the site is occupied by the building of the Faculty of Arts of the Ain Shams University (108)

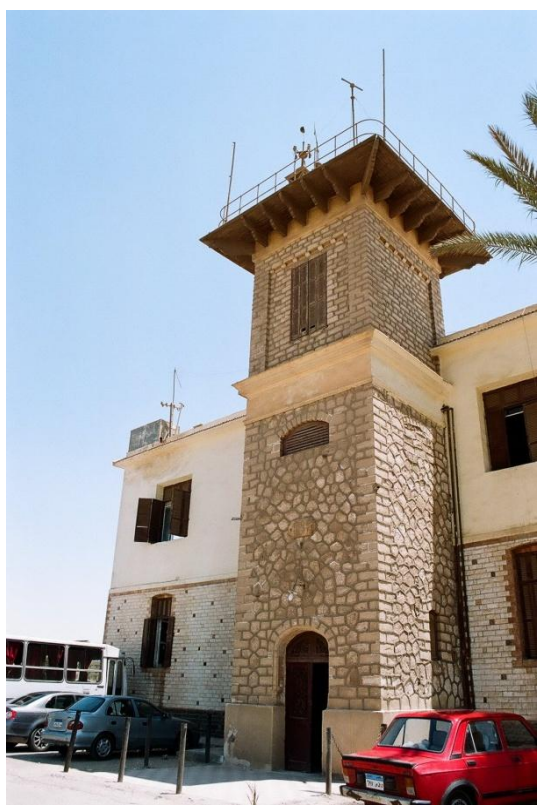

Figure 6: The administrative building at Helwan Observatory.

The platform atop has meteorological instruments installed.

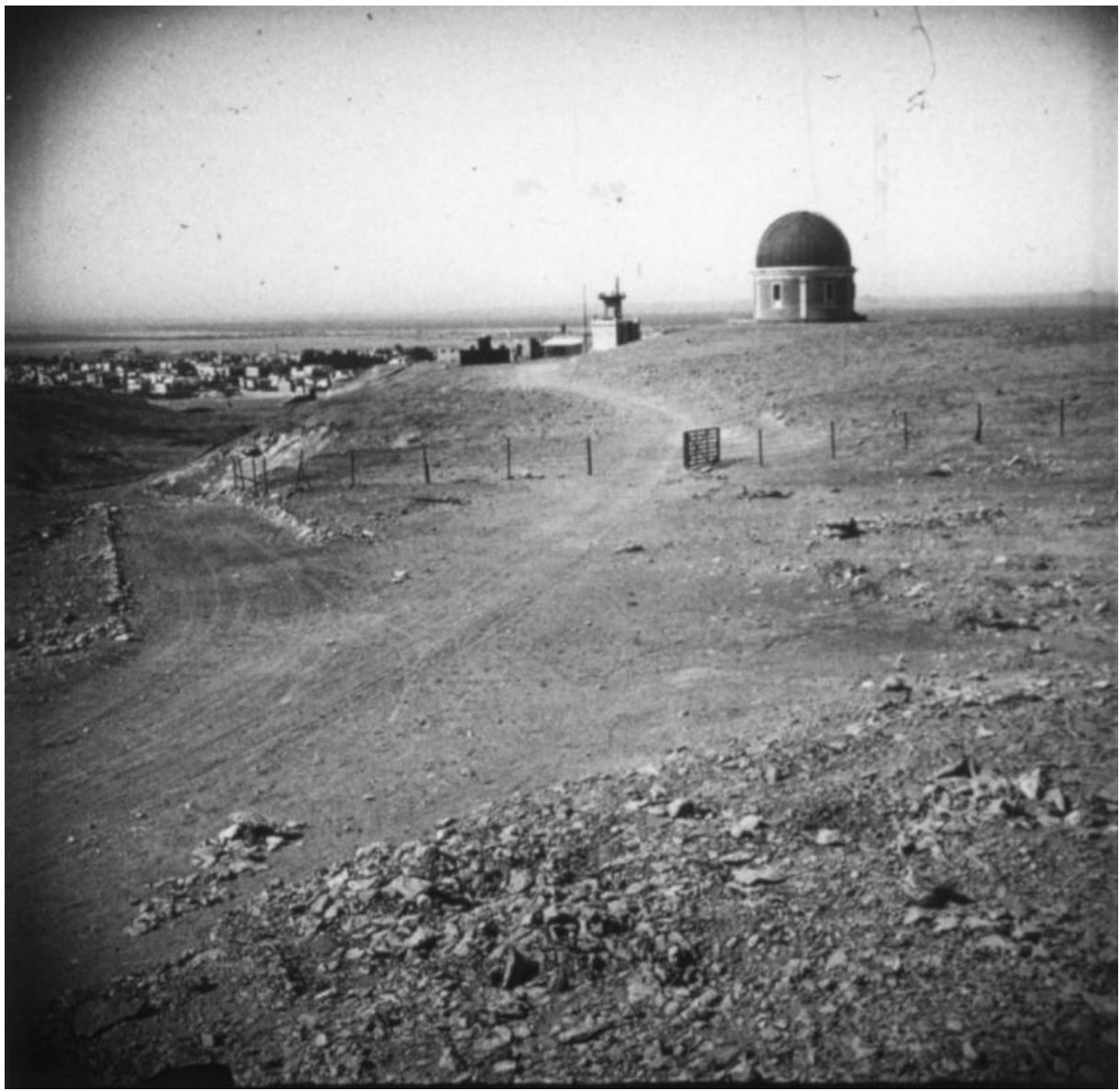

Figure 7: Site of the Helwan Observatory.

The dome is located near to the highest point on the site. To the left, at a lower elevation, is the tower of the administrative building shown in the previous Figure. To the left of that, in the valley below, is the town of Helwan. Beyond that is the Nile itself. A pyramid can just be made out on the horizon near to the right hand edge of the photograph.

Photograph by Harold Knox-Shaw (image courtesy of Peter Knox-Shaw and Anne Charles)

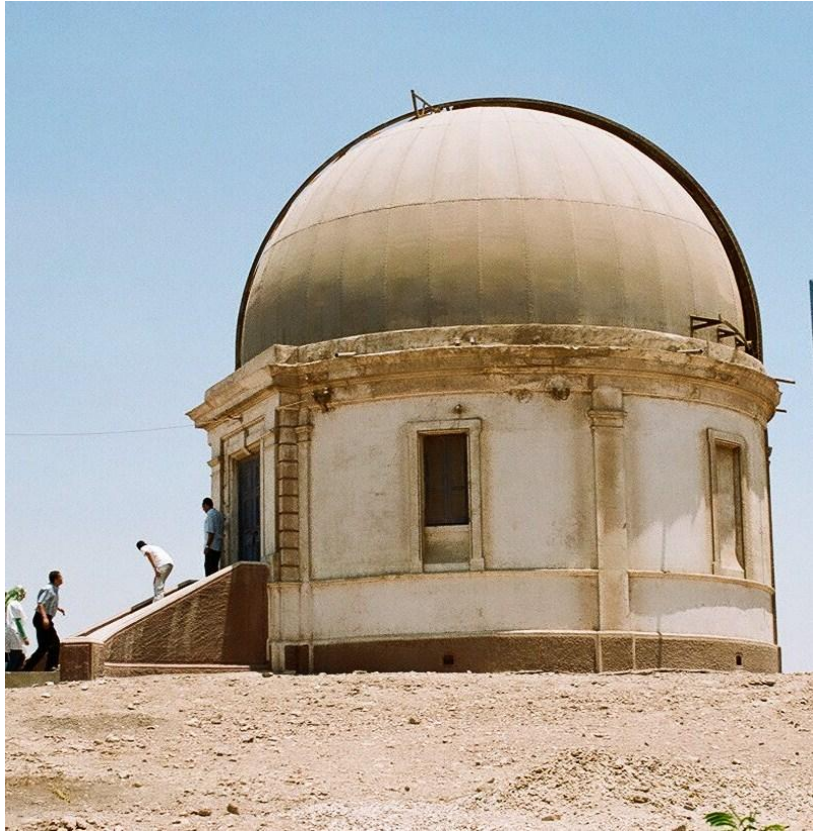

Figure 8: The dome of the Reynolds Telescope

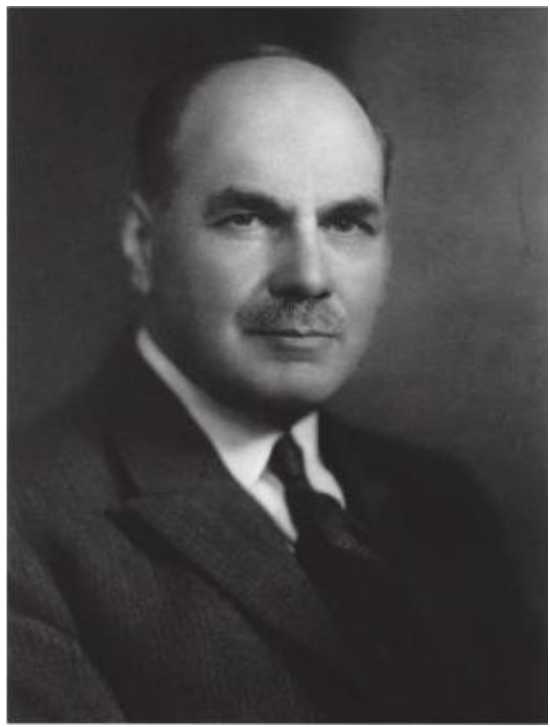

Figure 9: John Henry Reynolds (1874-1949)

(Royal Astronomical Society)

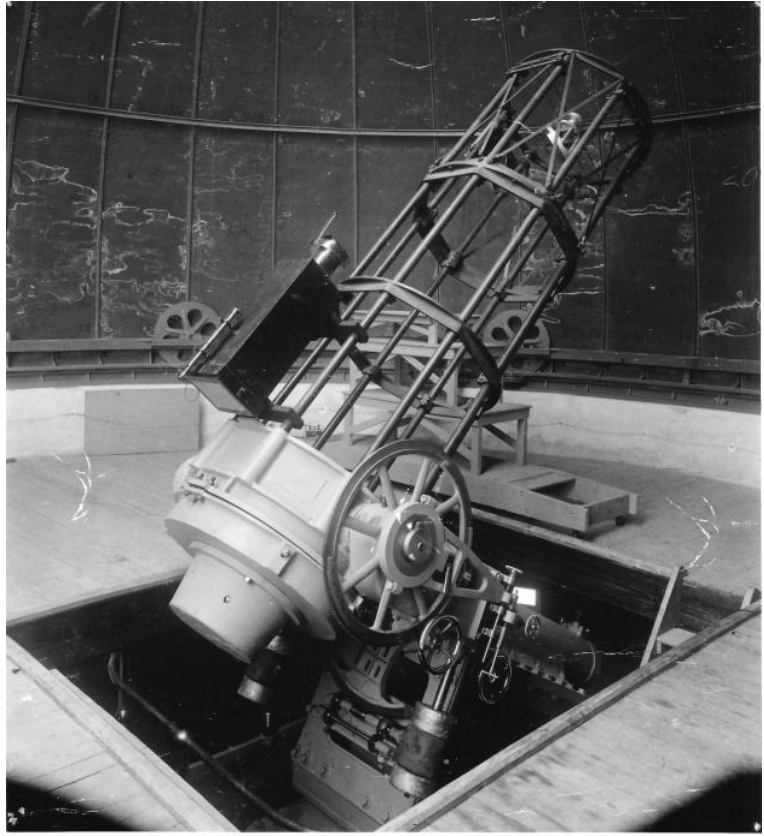

Figure 10: The 30-inch Reynolds Telescope at Helwan

(Royal Astronomical Society)

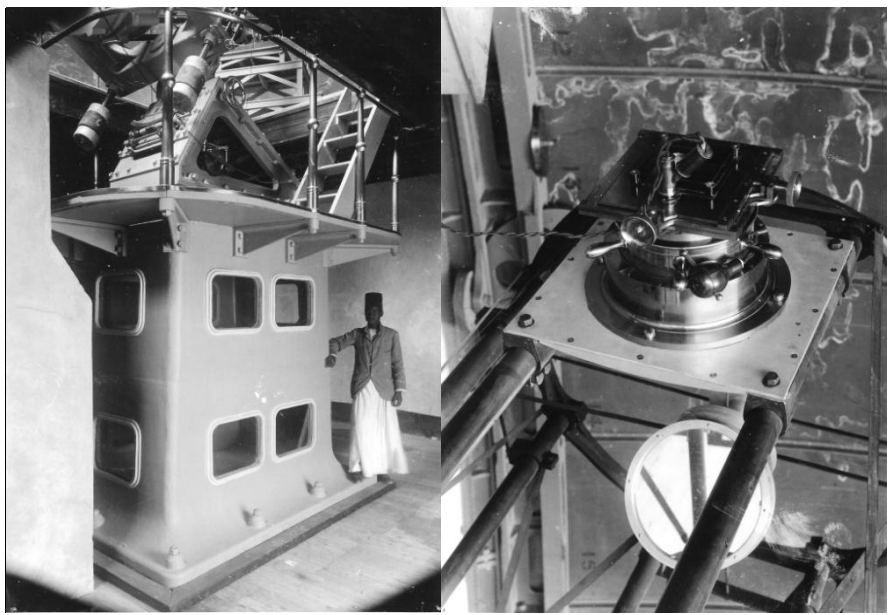

Figure 11: Details of the Reynolds Telescope

Left: the telescope pier, beneath the observatory floor. Right: the secondary mirror and the plate-holder (109).

(Royal Astronomical Society)

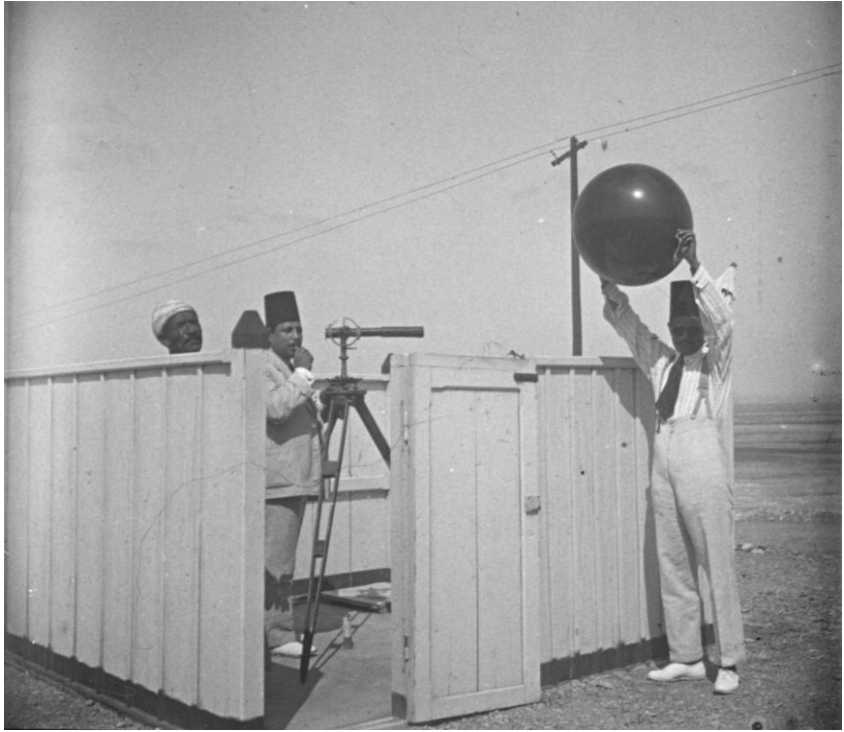

Figure 12: Preparing to launch a meteorological balloon from the Helwan site

Photograph by Harold Knox-Shaw (image courtesy of Peter Knox-Shaw and Anne Charles)

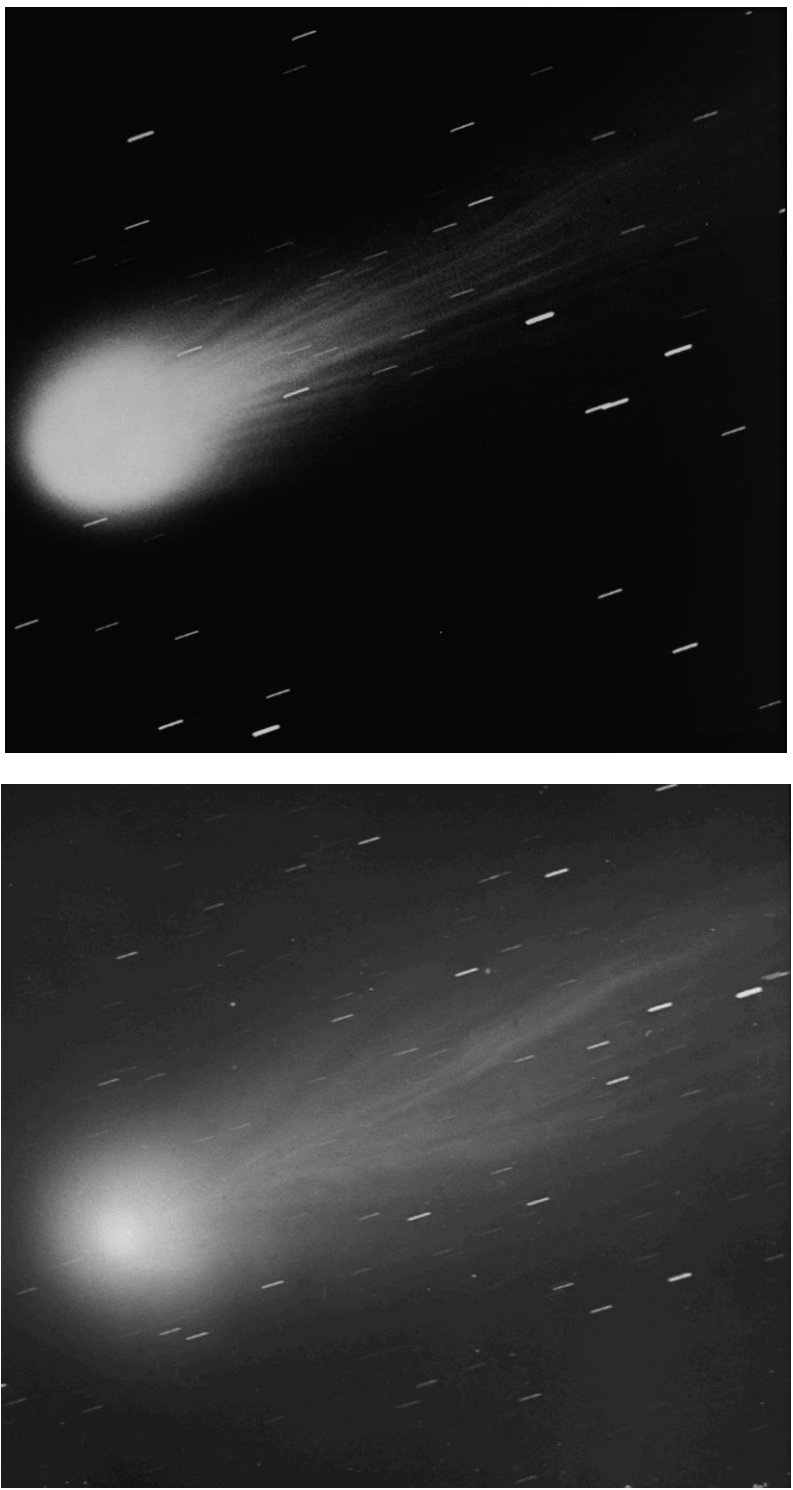

Figure 13: Halley's Comet in 1910

Top: 29 May. Bottom: 2 June

(photographs by H. Knox-Shaw, used with permission of Peter Knox-Shaw and the Science & Society Picture Library)

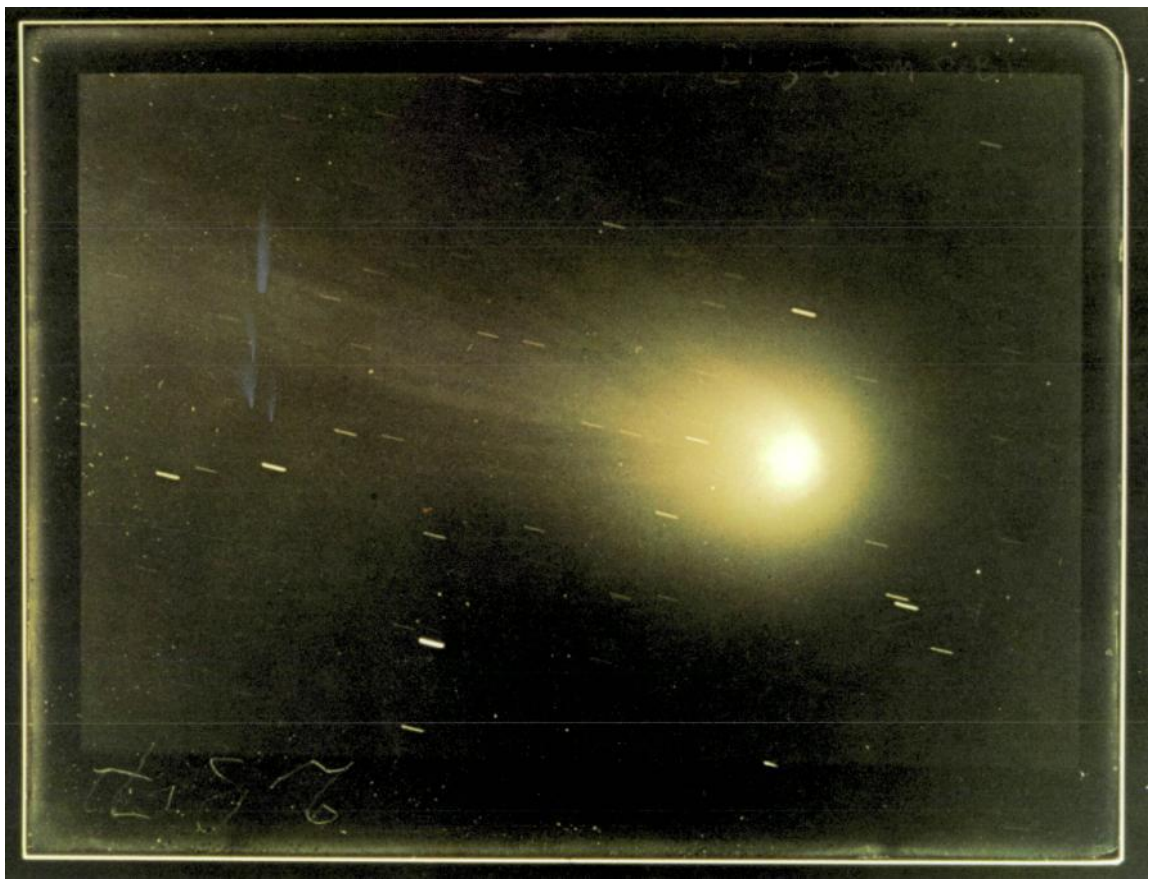

(a)

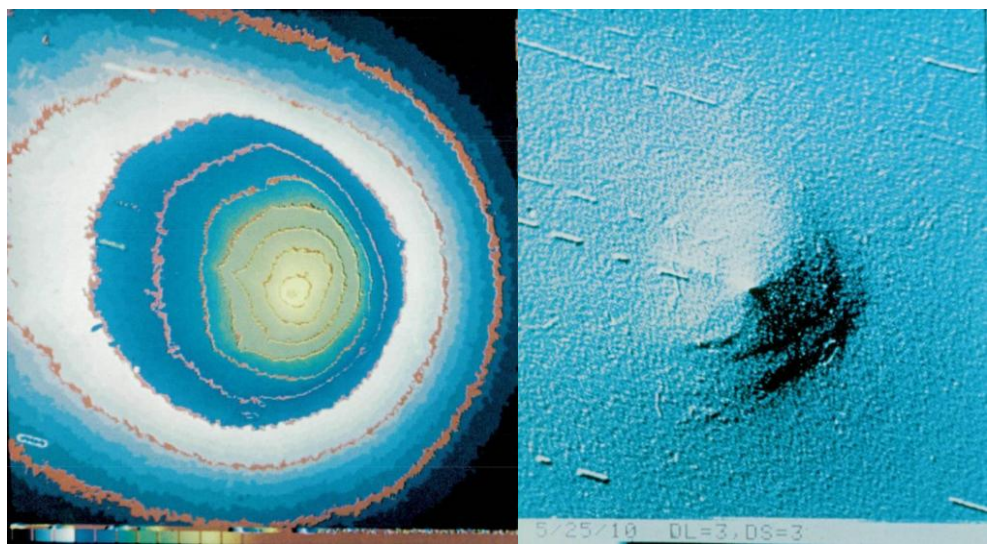

(b)                                    (c)

Figure 14: Halley's Comet on 25 May 1910

(a) Print of one of Knox-Shaw's original plates. (b) Brightness gradient map of the region near the nucleus. (c) Processed image. A small anti-solar projection can be see projecting left and above horizontal from the nucleus in (b) and (c) (Images courtesy of Dan Klinglesmith III (35))

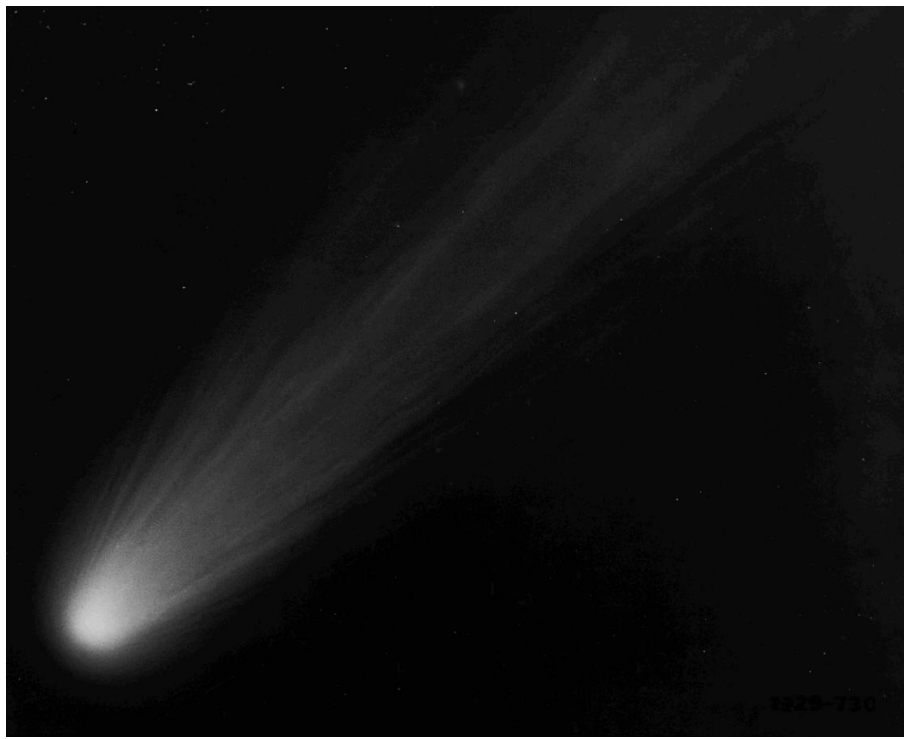

Figure 15: Comet 19P/Borelly. 28 October 1911, 15 min exposure (110)

(photograph by H. Knox-Shaw, used with permission of Peter Knox-Shaw and the Science & Society Picture Library)

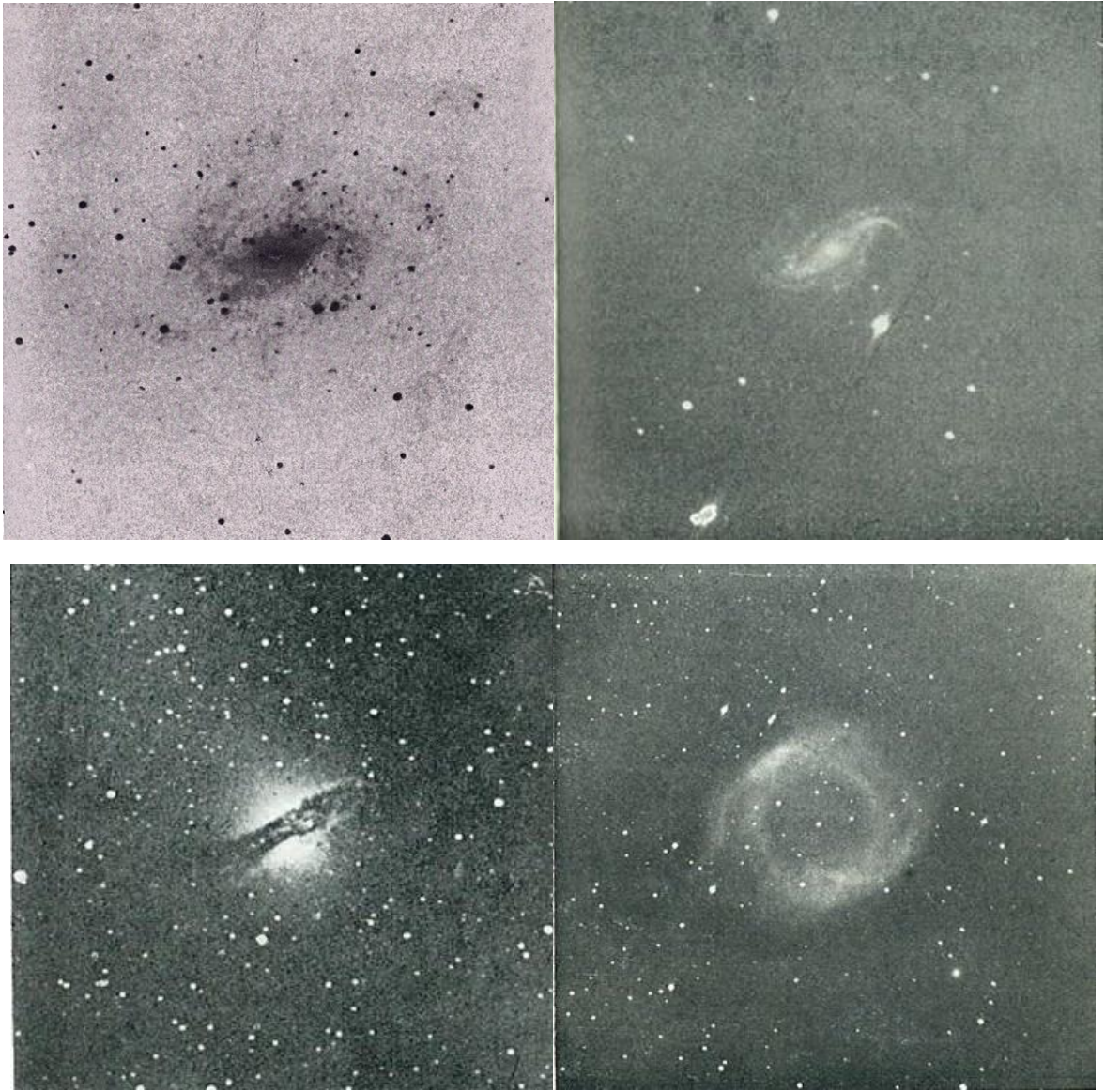

Figure 16: Photographs of nebulae taken by Knox-Shaw with the Reynolds Telescope

Clockwise from top left: NGC 300 (Sd galaxy in Sculptor), NGC 613 (SB galaxy in Sculptor), NGC 5128 (Centaurus A), NGC 7293 (planetary nebula in Aquarius)

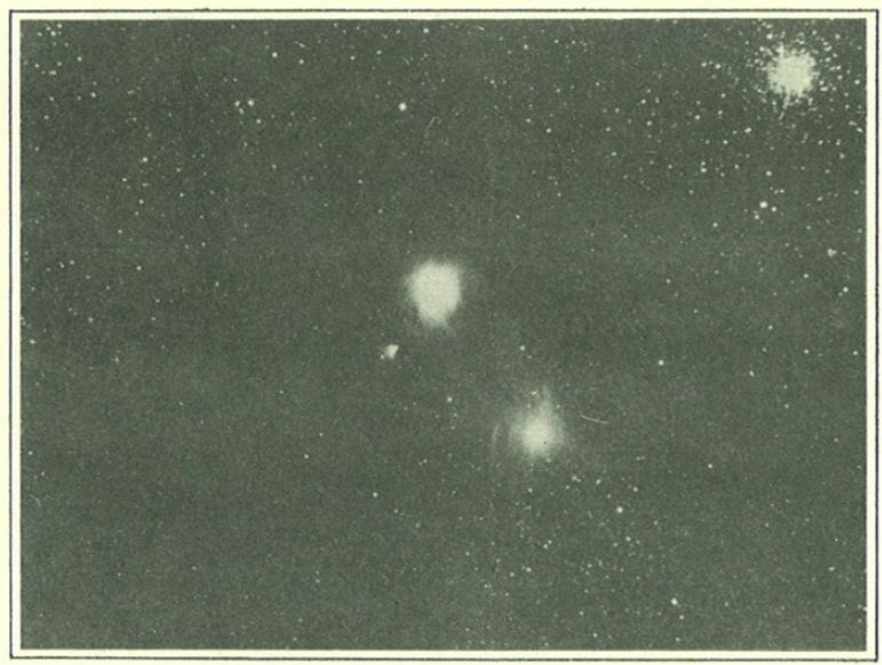

Figure 17: The region of the Variable Nebula in Corona Australis (NGC 6729)

(a) The bright nebulosity just above centre is NGC 6726-7. NGC 6729 is just to the lower left of this. The globular cluster at top right is NGC 6723. H. Knox-Shaw from ref. (111)

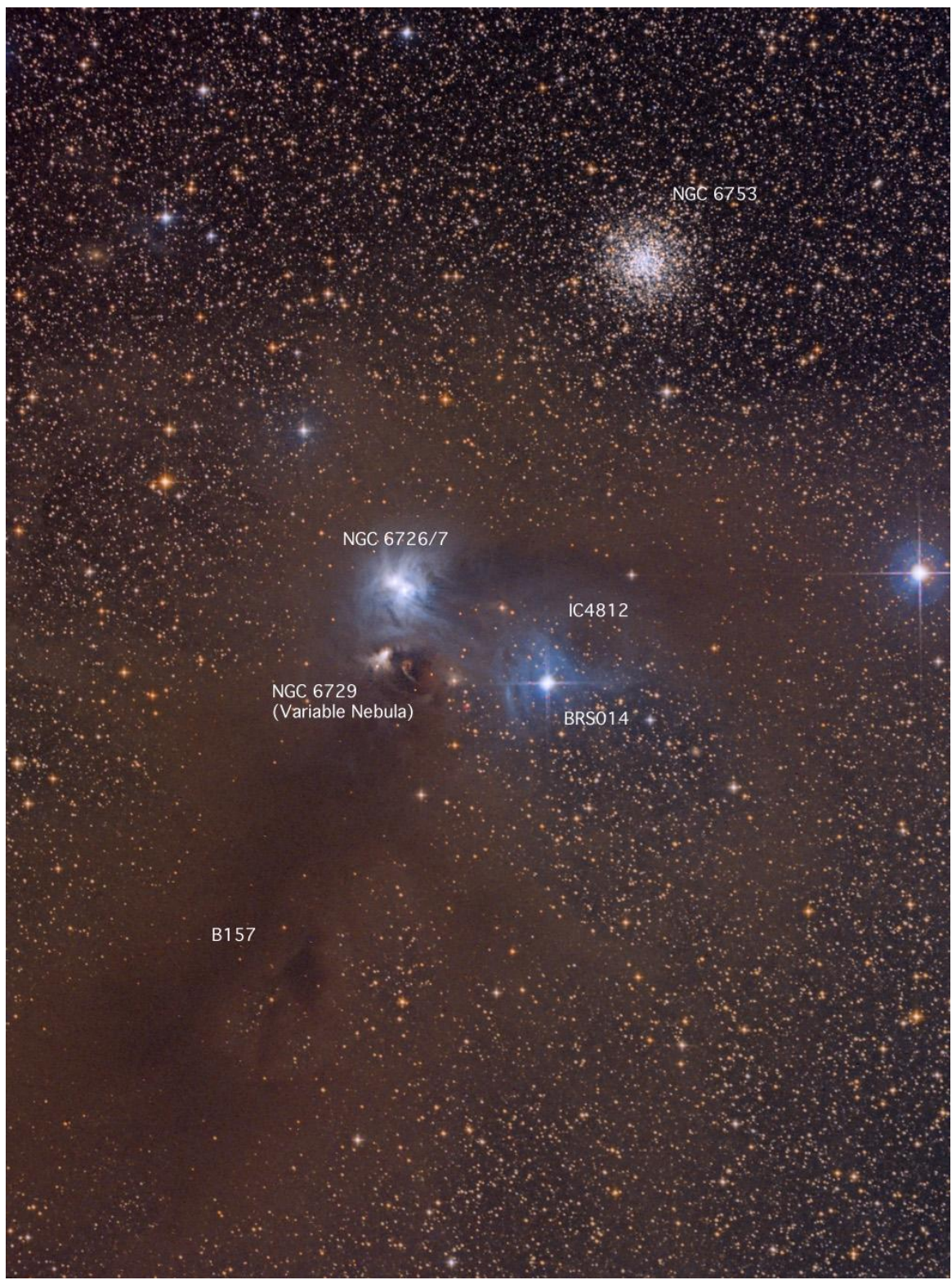

*(b)* Modern image of the region. Takahashi E210 astrograph, SBIG ST8300M CCD camera. Exposure 45 min total RGB (R 3x5min, G 3x5min, B 3x5min). Taken from Mersing, peninsula Malaysia by Remus Chua, Singapore *(note to Editor: I wonder if this image might be suitable for the Journal cover – I could arrange an unannotated version if required)*

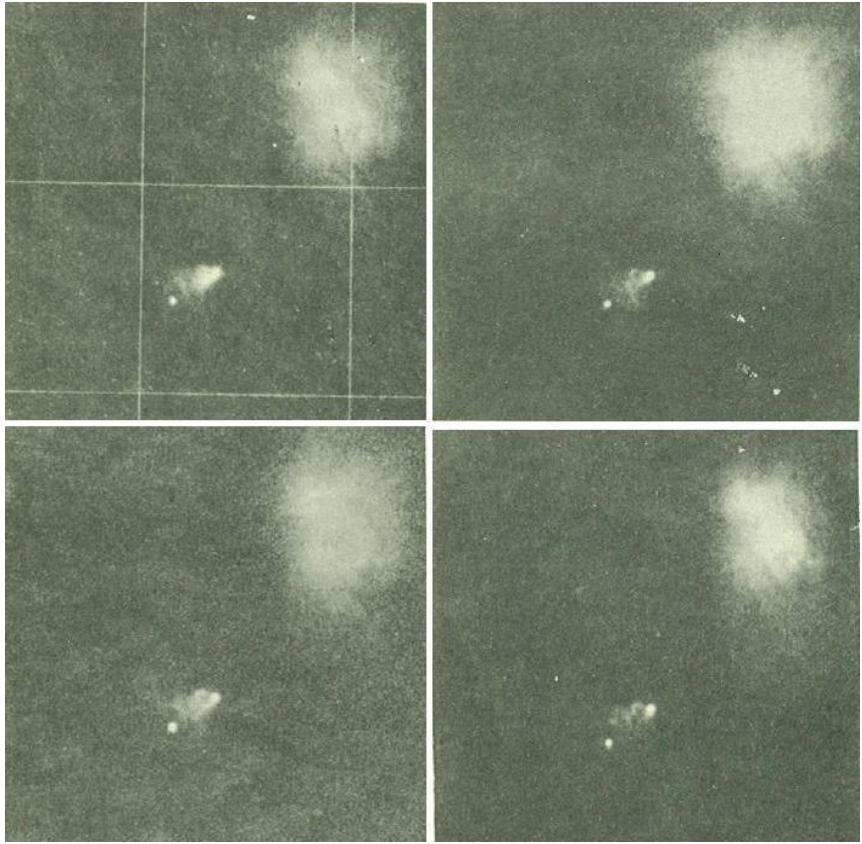

Figure 18: The Variable Nebula in Corona Australis (NGC 6729) in 4 different "states"

Clockwise from top left: Type A (7 July 1916), Type C (8 September 1915), Type E (26 July 1916) and Type G (6 August 1913). H. Knox-Shaw from ref. (111)

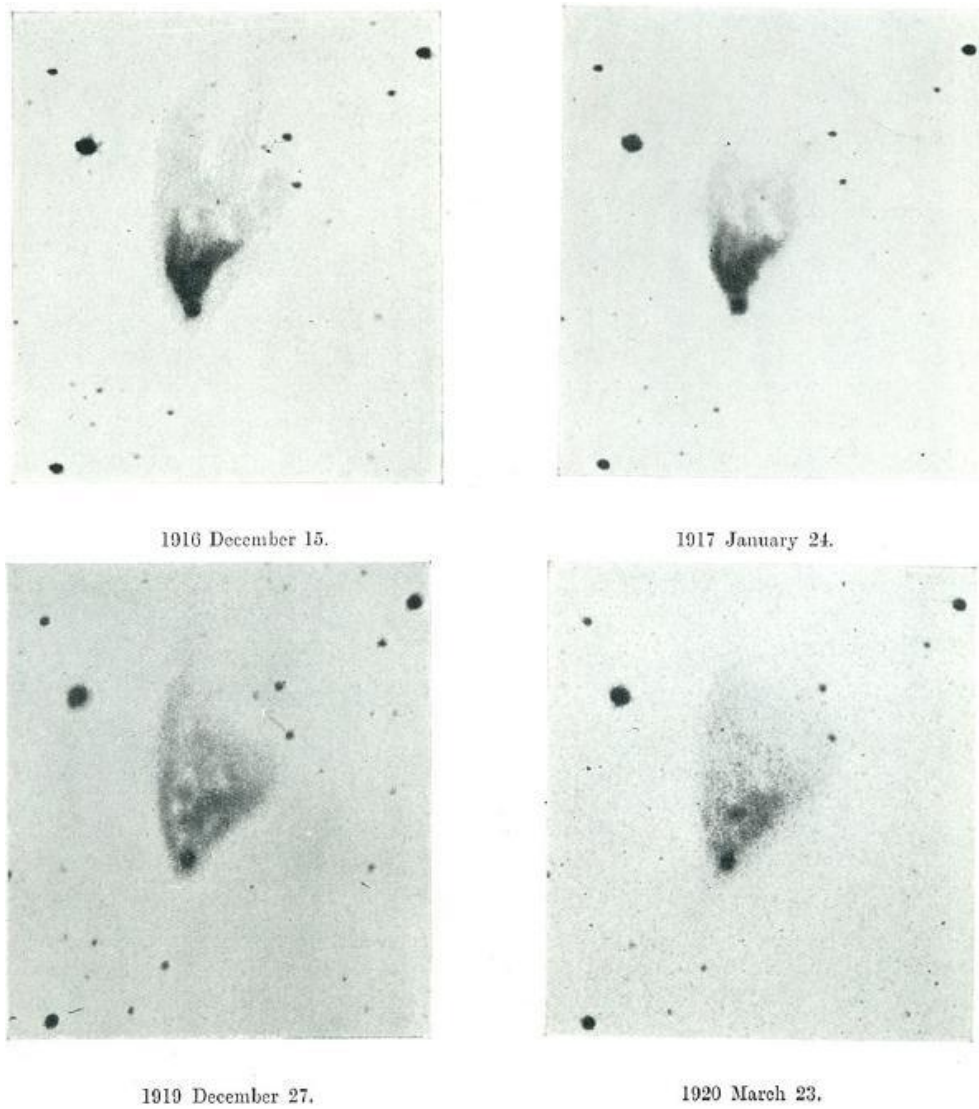

1916 December 15.

1917 January 24.

1919 December 27.

1920 March 23.

Figure 19: Hubble's Variable Nebula (NGC 2261). H. Knox-Shaw

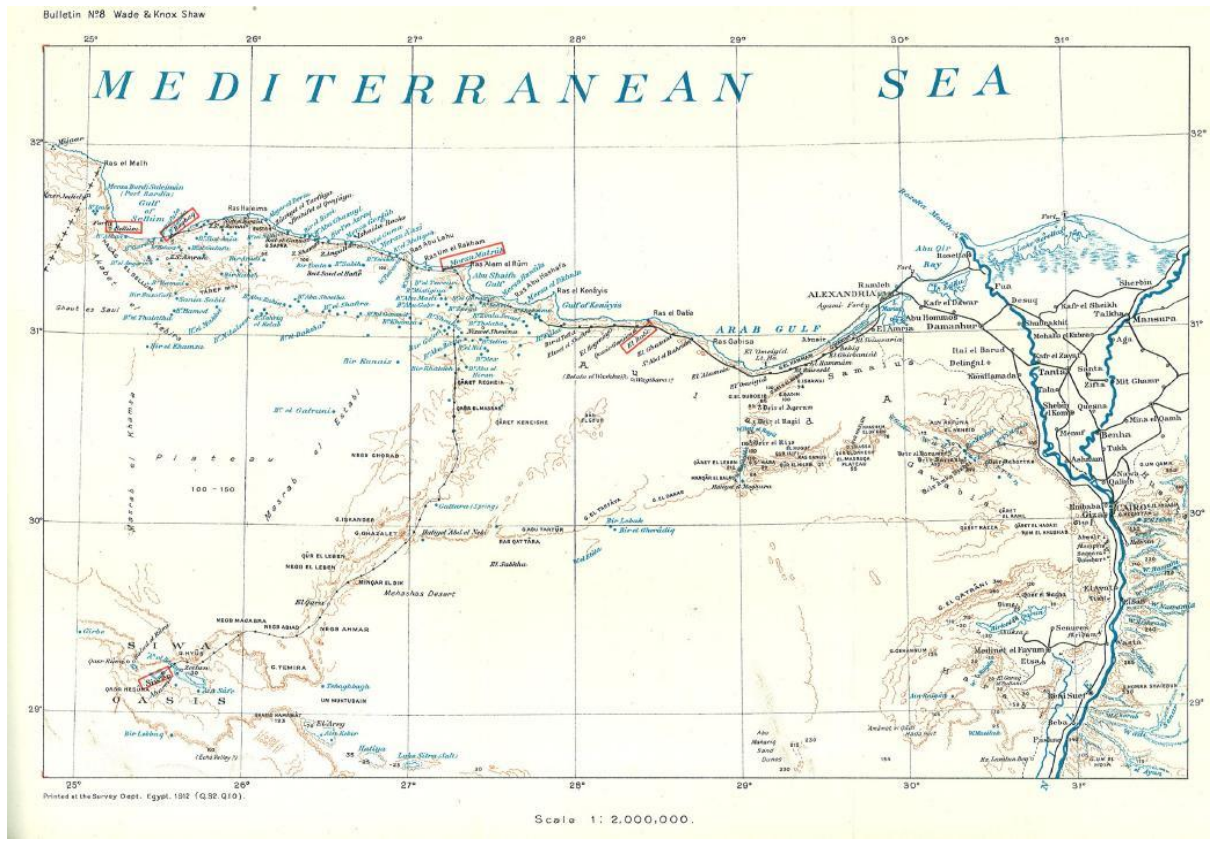

Figure 20: Map of northern Egypt and the Mediterranean coast

From reference (112)

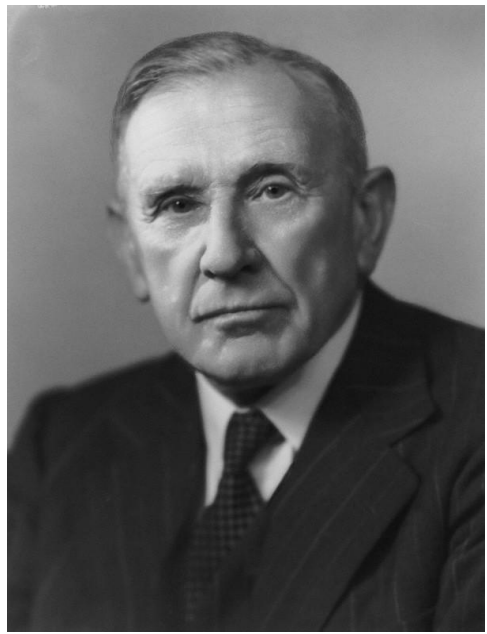

Figure 21: Harold Edwin Hurst (1880-1978)

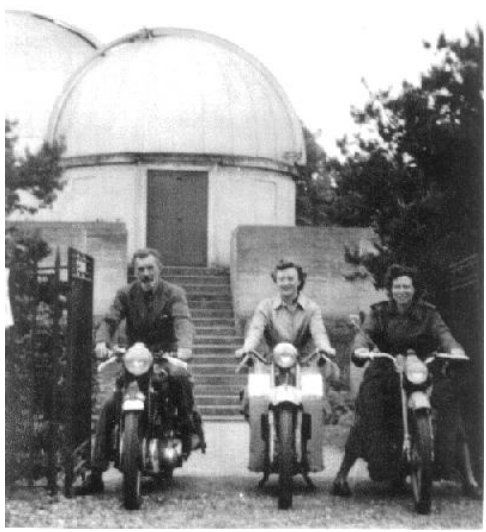

Figure 22: C.C.L. Gregory (1892-1964), left, with Margaret Peachey (later Margaret Burbidge) and A. Carew at the University of London Observatory, Mill Hill, ca. 1950

(image courtesy of Ian Howarth and the University of London Observatory)

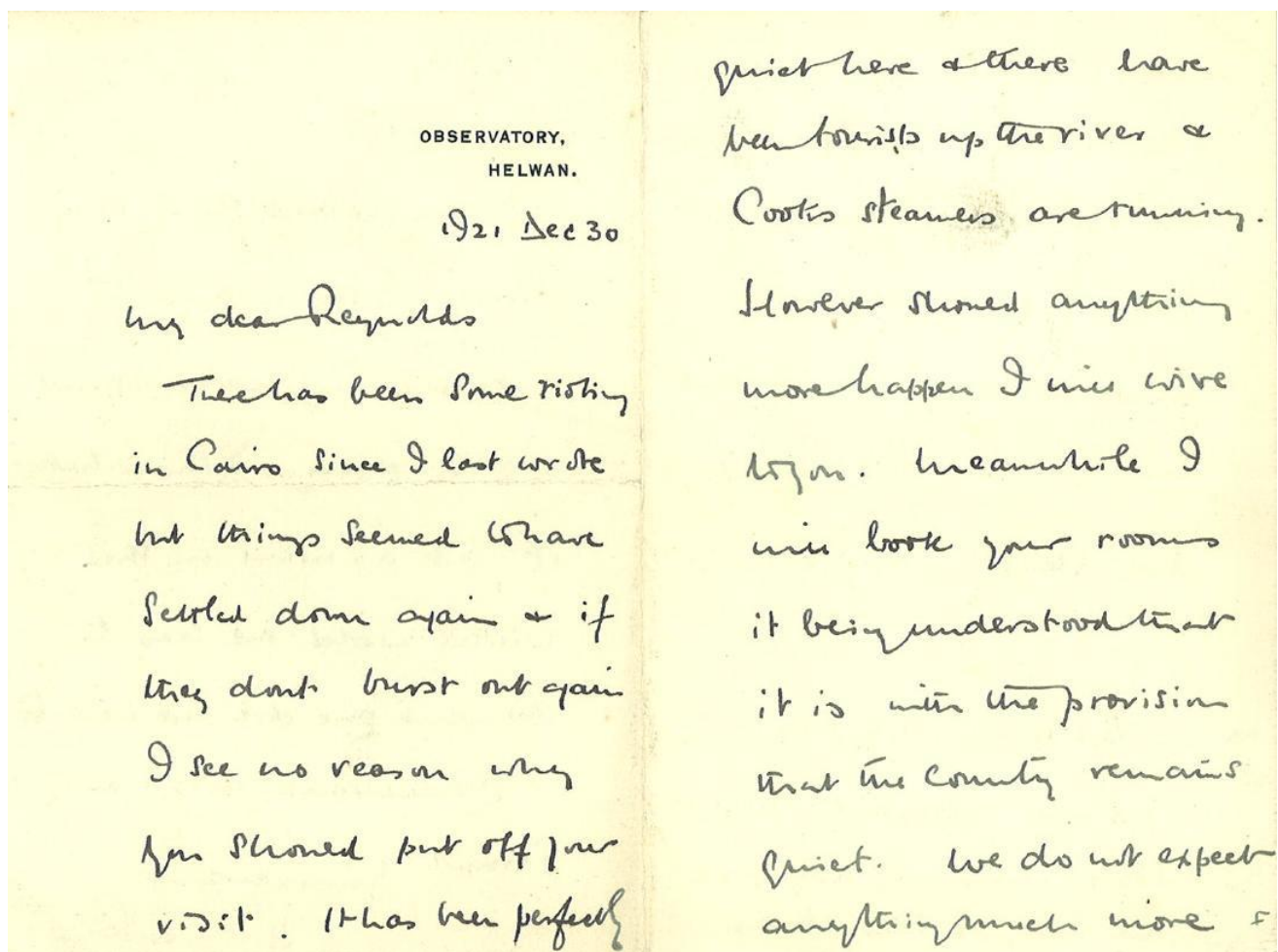

Figure 23: Letter from Knox-Shaw to Reynolds of 30 December 1921 referring to the situation in Cairo (Royal Astronomical Society)

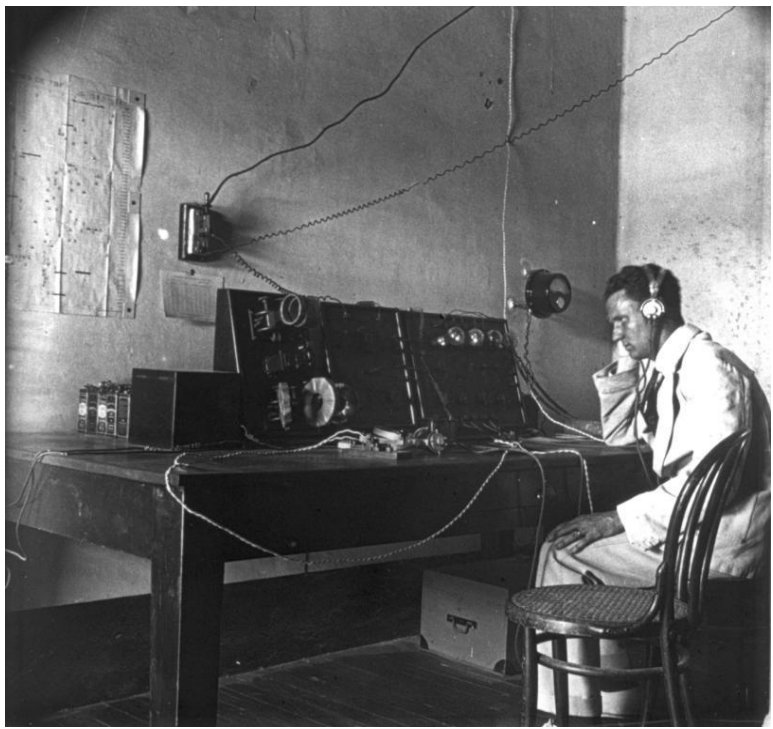

Figure 24: Burndept Ultra IV radio receiver (113)

Photograph by Harold Knox-Shaw (image courtesy of Peter Knox-Shaw and Anne Charles)

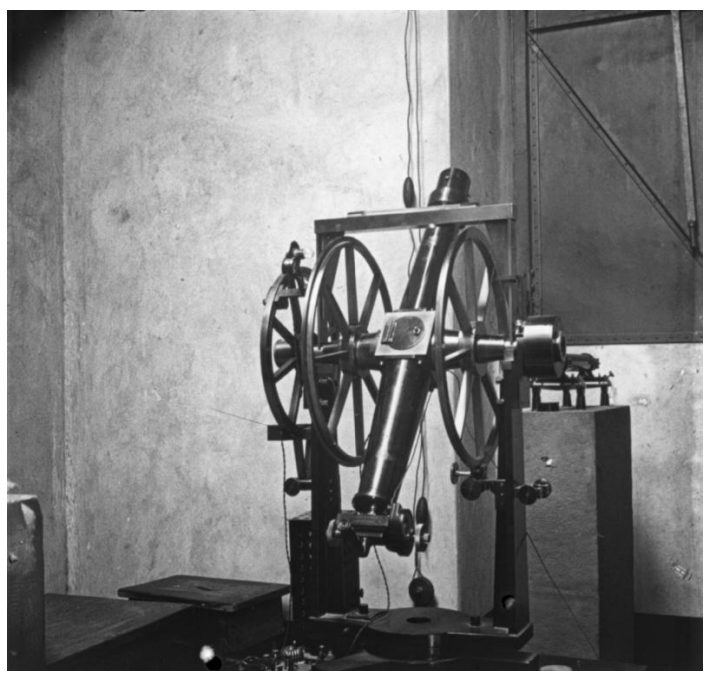

Figure 25: The Brunner Transit Circle. The telescope was housed in a separate building at Helwan. The instrument had an aperture of 55 mm and a focal length of 750 mm. It contained an electrically illuminated travelling wire micrometer by Cooke. The switch at the bottom of the photograph was used to record transit times on the observatory's chronometer.

Photograph by Harold Knox-Shaw (image courtesy of Peter Knox-Shaw and Anne Charles)

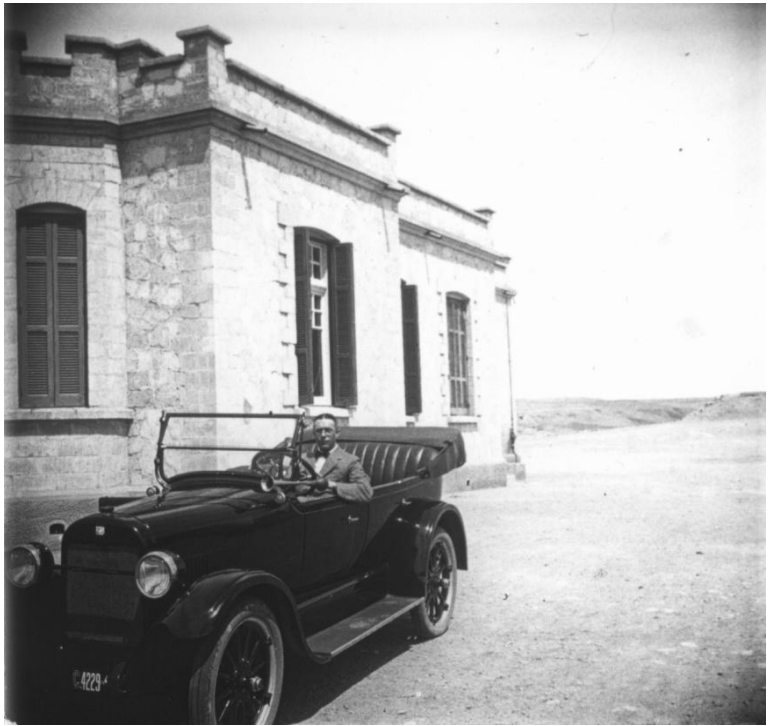

Figure 26: Harold Knox-Shaw in a Buick motor car (114)

(image courtesy of Peter Knox-Shaw and Anne Charles)

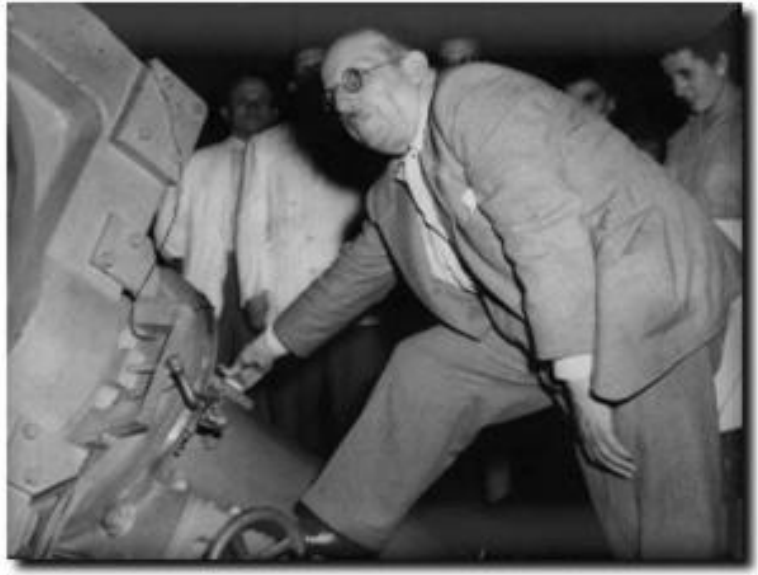

Figure 27: Mohammed Reda Madwar (1893-1973) next to the Reynolds telescope

(Archives of the National Research Institute of Astronomy and Geophysics, Egypt)